\documentclass{article}
\usepackage{t1enc}
\usepackage[utf8]{inputenc}
\usepackage[english]{babel}
\usepackage{graphicx}
\usepackage{amssymb}
\usepackage{amsmath}
\usepackage{amsthm}

\def\OA{{\cal A}}
\def\BO{{\cal B}}

\def\HS{{\cal H}}
\def\CONF{{\cal C}}
\def\FA{{\cal F}}

\def\SDK{{\cal K}}
\def\FMK{{\cal L}}
\def\ST{{\cal M}}
\def\CS{{\cal S}}
\def\PR{{\cal P}}
\def\vNA{{\cal N}}
\def\CO{{\mathbb C}}
\def\RE{{\mathbb R}} 
\def\IN{{\mathbb Z}}

\def\fel{{\frac{1}{2}}}

\def\UN{\mathbf{1}}
\def\OP{{\mathcal T}}
\def\qed{\ \vrule height 5pt width 5pt depth 0pt}
\def\cros{\raise1.9pt\hbox{$\scriptscriptstyle
          >$}\!\raise1.5pt\hbox{$\scriptstyle\triangleleft\,$}}
\theoremstyle{definition}\newtheorem{D}{Definition}
\theoremstyle{definition}
\theoremstyle{definition}\newtheorem{Prop}{Proposition}
\theoremstyle{definition}

\title{\bf On the concept of Bell's local causality \\ in local classical and quantum theory}
\author{\textit{G\'abor Hofer-Szab\'o}\thanks{Research Center for the Humanities, Budapest, email: szabo.gabor@btk.mta.hu} \\
\textit{P\'eter Vecserny\'es}\thanks{Wigner Research Centre for Physics, Budapest, email: vecsernyes.peter@wigner.mta.hu}}
\date{ }

\begin{document}
\maketitle

\begin{abstract}

The aim of this paper is to give a sharp definition of Bell's notion of local causality. To this end, first we unfold a framework, called local physical theory, integrating probabilistic and spatiotemporal concepts. Formulating local causality within this framework and classifying local physical theories by whether they obey local primitive causality --- a property rendering the dynamics of the theory causal, we then investigate what is needed for a local physical theory, with or without local primitive causality, to be locally causal. Finally, comparing Bell's local causality with the Common Cause Principles and relating both to the Bell inequalities we find a nice parallelism: Bell inequalities cannot be derived neither from local causality nor from a common cause unless the local physical theory is classical or the common cause is commuting, respectively.
\vspace{0.1in}

\end{abstract}

\section{Introduction}

In the history of causation spatiotemporal considerations always played an eminent role: they governed the general discourse in philosophy and informed the concrete theory constructions in physics. Just recall Hume's ideas on the contiguity of cause and effect, Newton's struggling with the action at a distance in his theory of gravitation, or Faraday's field theoretical program in electromagnetism. There is, however, an important milestone in the history of local causality, namely John Stewart Bell. Bell's merit is that he was able to translate the philosophical intuitions lying behind local causality into easily tractable mathematical terms which then set the scene for a whole research program in the foundations of quantum theory.

What are these philosophical intuitions? In a 1988 interview Bell formulates them as follows:  
\begin{quote}
``[Local causality] is the idea that what you do has consequences only   nearby, and that any consequences at a distant place will be weaker and will arrive there only after the time permitted by the velocity of light. Locality is the idea that consequences propagate continuously, that they don’t leap over distances.'' (Mann and Crease, 1988)
\end{quote}
Bell has returned to this intuitive notion of local causality from time to time and presented a more and more refined formulation of it. His line of reasoning, however, remained the same. Local causality excludes causal processes propagating faster than the speed of light but does not exclude correlations between spatially separated events. Such correlations, namely, can be brought about by a common cause operating in the past of the events in question. However, fixing the past of an event in a detailed enough manner, the state of this event in a locally causal theory will be fixed once and for all, and no other spatially separated event can contribute to it.

Looking at purely the logical structure of Bell's formulation of local causality, one can well see that it is an \textit{inference pattern from spatiotemporal to probabilistic relations}: if events are localized in the spacetime in a certain way, then they are to satisfy certain probabilistic independencies. Be these inferences as intuitive and applicable in the concrete physical praxis as they are, for a clear treatment something more is needed: a conceptual-formal \textit{framework} integrating spatiotemporal and probabilistic concepts in a common schema. Without such a framework, one could not account for the inferences from relations between spacetime regions to probabilistic independencies between, say, random variables. Where to find such a framework?

The most elaborate formalism used in physics offering a general method to connect spatiotemporal and probabilistic entities is quantum field theory, or its algebraic-axiomatic form, algebraic quantum field theory (AQFT) \textit{aka} local quantum physics (Haag, 1992). AQFT is a mathematically transparent theory ideal for analyzing various concepts related to local causality, such as the Bell inequalities (Summers, 1987a,b; Summers and Werner, 1988; Halvorson 2007); relativistic causality (Butterfield 1995, 2007; Earman 2014; Earman and Valente, 2014); or the closely related (see below) Common Cause Principle (Rédei 1997; Rédei and Summers 2002; Hofer-Szabó and Vecsernyés 2012a, 2013a). To our ends, however, the full formalism of AQFT would be too much. Our intention is simply to provide a \textit{minimal} framework which is needed to formulate Bell's notion of local causality in a strict fashion. We will call such a framework a \textit{local physical theory}. A local physical theory is a formal 
structure integrating the two most important components of a general physical theory: a spacetime structure and an algebraic-probabilistic structure. By using only few axioms in charactering local physical theories, our ambition is to cover as many concrete physical theories with spatiotemporal connotations as possible. Having a firm formal framework in hand, we can accomplish our primary goal which is to define Bell's notion of local causality in a clear-cut way and to relate it to other causality and locality concepts.

The paper is structured as follows. In Section 2 we set the mathematical framework of a local physical theory and spend some time to motivate the application of von Neumann algebras in this framework. Section 3 is devoted to the important concepts leading to causal dynamics of the observables in local physical theories, namely primitive causality and local primitive causality. In Section 4 we list and analyze further relativistic causality principles used in a local physical theory, such as parameter and outcome independence, local determinism and stochastic Einstein locality. In Section 5 we present Bell's own formulation of local causality and redefine it in the framework of local classical or quantum theories. In the same section we prove that local primitive causality makes a local physical theory to be locally causal. In Section 6 we relate local causality to causal stochastic dynamics in local classical theories without primitive causality. In Section 7 we compare local causality with the Common Cause Principle and relate both concepts to the Bell inequalities. We sum up in Section 8.

Our paper is fitting into a recent research line on a deeper conceptual and formal understanding of Bell's notion of local causality. Travis Norsen illuminating paper on local causality (Norsen, 2011) or its relation to Jarrett's completeness criterion (Norsen, 2009); the paper of Seevinck and Uffink (2011) aiming at providing a 'sharp and clean' formulation of local causality; or Henson's (2013b) paper on the relation between separability and the Bell inequalities all attest this renewed interest in local causality. We will comment on the points of contact with these papers underway. For a more philosopher-friendly and less technical version of our paper see (Hofer-Szabó and Vecsernyés 2014).

\section{What is a local physical theory?}

Let us start our project by defining a general framework, called local physical theory, which enables us to treat spatiotemporal and probabilistic entities in a common formalism. Instead of jumping directly to the full-fledged definition, we will proceed here  'inductively' by unfolding the notion of a local physical theory and specifying its different characteristic features step by step. Having listed these features we formulate the exact definition only at the end of the section. 

The central idea of a local physical theory is the \textit{association of local operator algebras to spacetime regions} regulated by the following physically motivated requirements (Haag, 1992):
\begin{enumerate}
\item \textit{Isotony.} Let $\ST$ be a globally hyperbolic spacetime\footnote{By a spacetime we mean a connected time-oriented Lorentzian manifold. A spacetime $\ST$ is called globally hyperbolic if $\ST$ contains a Cauchy hypersurface, which is by definition a subset $\CS\subset\ST$ such that each inextendible timelike curve in $\ST$ meets $\CS$ at exactly one point. (See (Pf\"affle, 2009) and references therein.)} and let $\SDK$ be a covering collection\footnote{For all $x\in\ST$ there exists $V\in\SDK$ such that $x\in V$.} of bounded, globally hyperbolic subspacetime regions of $\ST$ such that $(\SDK,\subseteq)$ is a directed poset under inclusion $\subseteq$. The net of local observables is given by the isotone map $\SDK\ni V\mapsto\OA(V)$ to unital $C^*$-algebras, that is $V_1 \subseteq V_2$ implies that $\OA(V_1)$ is a unital $C^*$-subalgebra of $\OA(V_2)$. The \textit{quasilocal algebra} $\OA$ is defined to be the inductive limit $C^*$-algebra of the net $\{\OA(V),V\in\SDK\}$ of local
$C^*$-algebras.\footnote{This formulation is a special case of the general category theoretical formulation of AQFTs in curved backgrounds (Brunetti and Fredenhagen, 2009). Namely, the functor from globally hyperbolic spacetimes to unital $C^*$-algebras is restricted to the full subcategory induced by the object $\ST$ and the (sub)collection $\SDK$ of its subobjects.}  

Sometimes \textit{additivity}, which is a stronger property than isotony, is also required for the net of observables: $\OA(V_1)\vee\OA(V_2)=\OA(V_1\cup V_2); V_1,V_2,V_1\cup V_2\in\SDK$, where $\vee$ refers to the generated algebra in $\OA$.

\item \textit{Microcausality} (also called as\textit{ Einstein causality}) is the requirement that $\OA(V')'\cap\OA \supseteq \OA(V),V\in\SDK$, where primes denote spacelike complement and algebra commutant, respectively.

\item \textit{$\mathcal{P}_\SDK$-covariance.} Let $\mathcal{P}_\SDK$ be the subgroup of the group $\mathcal{P}$ of global isometries of $\ST$ leaving the collection $\SDK$ invariant. A group homomorphism $\alpha\colon\mathcal{P}_\SDK\to\textrm{Aut}\,\OA$ is given such that the automorphisms $\alpha_g,g\in\mathcal{P}_\SDK$ of $\OA$ act covariantly on the observable net: $\alpha_g(\OA(V))=\OA(g\cdot V), V\in\SDK$.
\end{enumerate}
Here the possible spacetimes spread from Minkowski spacetime through stationary spacetimes to generic globally hyperbolic ones where no global Killing vector field exists. Choosing the collection $\SDK$ in a way that every $V\in\SDK$ contains only a finite number of elements of $\SDK$, one can consider local theories with locally finite degrees of freedom when the local algebras are finite dimensional. Otherwise the local algebras themselves are infinite dimensional.  

We would like to treat classical and quantum theories on an equal footing as far as possible. The difference between the two is that the quasilocal algebra of a local classical theory is required to be commutative while that of a local quantum theory  is required to be noncommutative. Thus, microcausality fulfils trivially in local classical theories. On the other hand, in local quantum theories it is usually required that the quasilocal algebra is `highly noncommutative' and the local algebras are `fat enough'. This is assured by algebraic Haag duality which is a stronger requirement than microcausality:
\begin{enumerate}
\item[4.Q] \textit{Algebraic Haag duality.} 
$\OA(V')'\cap\OA = \OA(V),V\in\SDK$. 
\end{enumerate}
Clearly, Haag duality is inherently connected to the noncommutativity of the observable algebra. In case of commutative local algebras Haag duality would imply that $\OA(V)=\OA$ for any $V\in\SDK$, that is the net structure of local algebras would be completely lost. To avoid this trivial net structure in local classical theories, one requires less than Haag duality:
\begin{enumerate}
\item[4.C] \textit{Intersection property for spacelike separated regions.} The intersection property
\begin{equation}\label{intersection_prop}
\OA(V_1)\cap\OA(V_2)=\OA(V_1\cap V_2);\qquad V_1,V_2,V_1\cap V_2\in\SDK 
\end{equation}
holds for spacelike separated regions $V_1,V_2\in\SDK$, that is $\OA(V_1)\cap\OA(V_2)=\OA(\emptyset):=\CO\,\UN_\OA$ for them.
\end{enumerate}
In case of local quantum theories this property follows from Haag duality and primitive causality (see below) if the net is additive and the quasilocal algebra is a factor, that is its center is trivial: $\OA'\cap\OA=\CO\,\UN_\OA$.\footnote{Let $V_1,V_2\in\SDK$ be spacelike separated regions. 
Due to Haag duality and additivity of the net 
\begin{equation}
\OA(V_1)\cap\OA(V_2)=\OA(V_1')'\cap\OA(V_2')'=(\OA(V_1')\vee\OA(V_2'))'
=\OA(V_1'\cup V_2')'. 
\end{equation}
Since $V_1'\cup V_2'$ always contains a Cauchy surface if $V_1$ and $V_2$ are spacelike separated bounded spacetime regions, we arrive at $\OA(V_1'\cup V_2')=\OA$ due to primitive causality. Therefore $\OA(V_1)\cap\OA(V_2)=\OA(V_1'\cup V_2')'=\OA'\cap\OA=:\mathrm{Center}\,\OA.$} We note that the intersection property (\ref{intersection_prop}) is not required for \textit{all} pairs $V_1,V_2\in\SDK$, since it would contradict to primitive causality which, as we will see, makes the dynamics to be deterministic.

Different physical realizations of a single local theory are given by unitary inequivalent representations $\pi\colon\OA\to\BO(\HS)$ of the quasilocal $C^*$-algebra $\OA$ by bounded operators $\BO(\HS)$ on a (separable) Hilbert space $\HS$. Inequivalent representations can be produced from essentially different states $\phi\colon\OA\to\CO$ through GNS--construction. Representations are required to be locally faithful not to loose local observables. Once a particular representation is chosen, one can consider the natural von Neumann algebra extension of the local algebras by taking weak closures $\vNA(V):=\pi(\OA(V))'', V\in\SDK$. 
\begin{enumerate}
\item[5.] \textit{Representation.} A locally faithful representation $\pi\colon\OA\to\BO(\HS)$ is chosen where a (strongly continuous) unitary representation $U\colon\PR_\SDK\to\BO(\HS)$ implements $\alpha\colon\PR_\SDK\to \textrm{Aut}\,\OA$. The local and quasilocal observables are extended as $\vNA(V):=\pi(\OA(V))'', V\in\SDK$ and $\OA_\HS:=\overline{\cup_{V\in\SDK}\vNA(V)}\subset\BO(\HS)$, respectively.
\end{enumerate}
It is easy to see that the net $\{\vNA(V),V\in\SDK\}$ of local von Neumann algebras given above also obeys \textit{isotony}, \textit{microcausality} in the sense that $\pi(\OA(V'))'\cap\BO(\HS)\supseteq \vNA(V),V\in\SDK$, and \textit{$\mathcal{P}_\SDK$-covariance}. Since we concentrate on local and causal properties we do not consider further requirements on the representation $\pi$, e.g. how a vacuum representation can be characterized and be chosen among the allowed representations.\footnote{However, to stay within the quasi-equivalence class of the representation $\pi$ one considers only states in the folium of $\pi$ (Haag, 1992), that is normal states of $\pi(\OA)''$ which lead to locally normal states, that is normal states by restricting them to the local von Neumann algebras $\vNA(V), V\in\SDK$.}  

Here we would like to briefly comment on the use of von Neumann algebras as local algebras in local classical theories. The crucial point is the link between von Neumann algebras and $\sigma$-algebras. Every element $S\subset\Omega$ of a $\sigma$-algebra $(\Omega,\Sigma)$ determines a projection $\chi_S$ in the abelian ${}^*$-algebra $\FA(\Omega,\CO)$ of complex functions on $\Omega$, namely, $\chi_S$ is the characteristic function of the subset $S\in\Sigma$. In general, we would like to translate local $\sigma$-algebras $(\Omega,\Sigma)$ to local commutative operator algebras generated by projections $\chi_S, S\in\Sigma$ in the function algebra $\FA(\Omega,\CO)$. This abundance of projections is, however, the reason why the local operator algebras cannot be represented by commutative $C^*$-algebras in a local classical theory. Namely, a commutative unital (nonunital) $C^*$-algebra, according to the Gelfand duality, is isomorphic to the algebra of complex valued continuous functions (vanishing at infinity) 
on a (locally) compact Hausdorff topological space. However, unless the topology is discrete, such algebras generally do not contain nontrivial projections at all. Therefore one is to consider commutative \textit{von Neumann algebras} in local classical theories as local operator algebras which are not only rich enough in projections, but also are generated by them.

The paradigmatic case of a commutative von Neumann algebras is the space of complex-valued essentially bounded measurable functions $L^\infty(\Omega,\Sigma,\mu)$ on the $\sigma$-finite measure space $(\Omega,\Sigma,\mu)$. This Neumann algebra is generated by the subclass $\{\chi_S, S\in \Sigma\}$ of characteristic functions on $\Omega$, and acts on the separable Hilbert space $L^2(\Omega,\Sigma,\mu)$ by multiplication. This subclass of characteristic functions, or equivalently, the sets of their supports form the $\sigma$-algebra $(\Omega,\Sigma)$ of classical events. The lattice operations and the algebra operations relate to one another as follows: $\chi_S\chi_T=\chi_{S\wedge T}, \chi_S+\chi_T-\chi_S\chi_T=\chi_{S\vee T}$. This $\sigma$-algebra, however, is not the most general $\sigma$-algebra one can imagine, since not every $\sigma$-algebra can be equipped by a $\sigma$-finite measure $\mu$. Nevertheless, they give us a rich enough set of examples for classical theories. The probability measure $p$ on the
corresponding $\sigma$-algebra $(\Omega,\Sigma)$ can be provided by any normal state $\omega$ on the  von Neumann algebra $L^\infty(\Omega,\Sigma,\mu)$ by $p_\omega(S):= \omega(\chi_S), S\in\Sigma$.

It is a further question as to what kind of local $\sigma$-algebras can correspond to local classical theories, e.g. to classical field theories with configuration space $F^{\ST}:=\{\Phi\colon\ST\to F\}$ with field values $F=\RE^n,\CO^n$, for example. The maximal $\sigma$-algebra of classical events one can imagine is $(F^{\ST},\PR(F^{\ST}))$ given by the power set $\PR(F^{\ST})$ of the set of field configurations. One needs also narrower $\sigma$-algebras in tune with the net structure of the theory. This is done by taking local equivalence classes of those configurations which have the same field values on a given region $V\in\SDK$. Two field configurations $\Phi,\Psi\in F^\ST$ are said to be \textit{locally $V$-equivalent}, $\Phi\sim_V\Psi$, if $\Phi_{\vert V}=\Psi_{\vert V}$. The isotone net structure $\{(F^{\ST},\Sigma(V)), V\in\SDK\}$ of unital $\sigma$-subalgebras $\Sigma(V)\subset\PR(F^{\ST})$  can be given by the `cylindrical subsets' of $F^{\ST}$ corresponding to the image sets of canonical projections
$Z_V\colon\PR(F^{\ST})\to\PR(F^{\ST}), V\in\SDK$, which map a set $S$ of configurations onto the corresponding union of $V$-equivalence classes of configurations in $S$: 
\begin{equation}\label{cylind_set}
\PR(F^{\ST})\ni S\mapsto Z_V(S):=\{ \Phi\in F^{\ST} \, | \exists \Psi\in S: 
\Phi_{\vert V}= \Psi_{\vert V}\}\in\Sigma(V):=Z_V(\PR(F^{\ST})).
\end{equation}
Clearly, the net $\{(F^{\ST},\Sigma(V)), V\in\SDK\}$ --- or $\{\Sigma(V), V\in\SDK\}$, for short --- is $\PR_\SDK$-covariant. The hard and unsolved problem is to give a probability measure on the $\sigma$-algebra $(F^{\ST},\PR(F^{\ST}))$ or on a meaningful $\sigma$-subalgebra of it. We can avoid this conundrum by choosing a locally finite covering of $\ST$, that is choosing a subnet $\SDK^m\subset\SDK$ in a way that every $V\in\SDK^m$ contains only a finite number of elements of $\SDK^m$, and restricting the field configurations to be piecewise constant on regions corresponding to minimal elements in $\SDK^m$.  The power set of this configuration space $F^{S^m}$, where $S^m$ denotes the set of minimal elements in $\SDK^m$, can also be mapped into local $\sigma$-algebras $(F^{S^m},\Sigma_m(V)), V\in\SDK^m$ as before in (\ref{cylind_set}). Although the maximal local $\sigma$-algebra $\Sigma_m(V^m)$ of a minimal region $V^m\in S^m$ is isomorphic to the power set $\PR(F)$ of field values, one can restrict them
to the Borel $\sigma$-subalgebra of $\PR(F)$. Then a generic local $\sigma$-algebra $\Sigma_m(V), V\in\SDK^m$ is isomorphic to a finite product of the copies of corresponding Borel $\sigma$-subalgebras, because $V$ is covered by a finite subset of $S^m$. We can simplify further the situation by restricting the field values $F$ to a finite set. In our example used below $F=\IN_2$, the group with two elements, represented by the integers $\pm 1$. In that case the local $\sigma$-algebra of a minimal region $V^m\in S^m$ is finite, $\Sigma_m(V^m)=\PR(\IN_2)$, hence the corresponding local von Neumann algebra is finite (two) dimensional, the two nontrivial projections correspond to the two nontrivial subsets of $\IN_2$.  

Last but not least, we would like to stress that the projections $\chi_S, S \in\Sigma(V)$ in the local von Neumann algebras do not possess a direct spacetime localization: they project to subsets of $F^{\ST}$ and not to those of $\ST$.
\vspace{0.1in}

\noindent
Inspired by the above considerations, we define a local physical theory as follows: 
\begin{D}\label{LPT} 
A \textit{local physical theory} (LPT) is a net $\{\vNA(V),V\in\SDK\}$ of local von Neumann algebras associated to a directed poset $\SDK$ of globally hyperbolic bounded regions of a globally hyperbolic spacetime $\ST$. The net  satisfies \textit{isotony}, \textit{microcausality},  \textit{$\mathcal{P}_\SDK$-covariance}, and  \textit{intersection property for spacelike separated regions}. If the local von Neumann algebras are commutative, we speak about a \textit{local classical theory} (LCT), if they are noncommutative, we speak about a \textit{local quantum theory} (LQT).
\end{D}
Our aim is to interpret and formulate Bell's notion of local causality in the framework of LPTs. Before turning to local causality, however, we need to understand what is a causal dynamics in a LPT and whether its existence is ensured by the very properties of a LPT. To this we turn in the next section.

\section{Causal dynamics} 

The motivation for causal dynamics (or causal time evolution) comes from classical field theory on a globally hyperbolic spacetime, where a global time parameter can be chosen. If the field equations of the theory are symmetric hyperbolic partial differential equations (see Geroch, 2010), then there exists an initial value formulation of the theory in the following form: given the initial values on (a piece of) a Cauchy surface, the time evolution equation provides a unique solution in the \textit{domain of dependence}\footnote{The domain of dependence $D(\CS)$ of a (piece of) a Cauchy surface $\CS$ consists of those points in $\ST$ for which any causal curve containing them intersects $\CS$.} of (that piece of) the Cauchy surface. This restriction of the complete influences of the initial values to the domain of dependence is that makes the dynamics of the theory \textit{causal}, since it forbids superluminal propagation (see Earman, 2014).

This causal dynamics has two basic properties: it is defined within a \textit{classical} theory, and it is \textit{deterministic} in the sense that fixing the (expectation) values of the observables at a certain time, the dynamics provides unique (expectation) values of the observables in the future or in the past (within the domain of dependence of the initial values). We will see that the properties of a LPT, classical or quantum, are not strong enough to provide us such a causal dynamics. An additional property, called \textit{primitive causality}, will ensure the dynamics in a LPT to be \textit{deterministic} in the above sense; and another, more restrictive property, called \textit{local primitive causality}, will ensure the dynamics in a LPT to be \textit{causal}. It will turn out that in the absence of primitive causality not only the causality of the dynamics (on the observables) is meaningless but also the notion of an initial state on the observables is missing. In this cas
 e a state on the quasilocal algebra
involves that one should prescribe the state on the proper Cauchy surface subalgebras for all time slices $t\in\RE$. Expectation values in a generic state of such LPTs are hardly expected to show any causal properties. However, at least in LCTs, one can restrict the set of possible states by sticking to states obtained by a special state extension procedure from an initial state on a single Cauchy surface subalgebra. This extension is special in the sense that it can be defined by a stochastic process obeying causal features. Hence, the extension procedure can be considered as a `dynamics on the states', and the causality of this dynamics, reflected in the causal properties of the expectation values, will arise from the causal properties of the underlying stochastic process. The rest of the section is devoted to what we mean by a causal dynamics on the observables or, in the absence of primitive causality, on the states, and how to ensure their 
existence in the framework of LPTs.
\vspace{0.1in}

\noindent
In case of stationary spacetimes, i.e. when a global timelike Killing vector field exists, a natural dynamics exists in LPTs on the observables, the \textit{covariant dynamics}: The one parameter isometry group $T\simeq (\RE,+)$ of $\ST$ generated by the global timelike Killing vector field leads to a one parameter automorphism group $\{\alpha_t,t\in T\}$ of the quasilocal observable algebra $\OA$ acting covariantly on the net (Requirement 3). In case of a generic globally hyperbolic spacetime $\ST$ no global timelike Killing vector field exists, therefore there is no natural dynamics  on the observables in LPTs. However, a foliation $\{\CS_t,t\in\RE\}$ of $\ST$ by Cauchy surfaces exists, which is indexed by a global time parameter. Such a foliation will lead to a dynamics on the observables  if the observable algebra corresponding to any of the Cauchy surfaces already exhausts the quasilocal observable algebra, that is \textit{primitive causality} holds:
\begin{enumerate}
\item[6.] \textit{Primitive causality.} For any covering collection $\SDK(\CS)\subseteq\SDK$ of any Cauchy surface $\CS$, one has $\OA_{\SDK(\CS)}=\OA$.
\end{enumerate}
The covering collection $\SDK(\CS_t)\subseteq\SDK$ of the Cauchy surface $\CS_t$ determines a subalgebra $\OA_{\SDK(\CS_t)}$ of $\OA_\HS$. Let us define the Cauchy surface algebra $\OA_{\CS_t}$ of $\CS_t$  by the injective limit algebra of a decreasing net of subalgebras corresponding to decreasing coverings (see (Brunetti and Fredenhagen, 2009) for details). Thus, in case of primitive causality any subalgebra $\OA_{\SDK(\CS)}$, hence any Cauchy surface subalgebra $\OA_\CS$ is equal to the whole quasilocal algebra $\OA$. Therefore, the injective algebra morphisms corresponding to embeddings of  globally hyperbolic Cauchy surface coverings into $\ST$ become isomorphisms and one obtains also algebra isomorphisms $\iota_t\colon\OA_{\CS_t}\to\OA, t\in\RE$ between the Cauchy surface algebras and the quasilocal algebra. Then the isomorphism $\alpha_{t',t}:=\iota_{t'}^{-1}\circ\iota_t\colon\OA_{\CS_t}\to\OA_{\CS_{t'}}$ provides the Cauchy time evolution isomorphism, that is the dynamics on 
the observables, between the Cauchy surface algebras corresponding to time slices $t$ and $t'$ in the chosen foliation. In the presence of a covariant dynamics the two dynamics coincide, $\alpha_{t',t}=\alpha_{t'-t}$ if the chosen foliation of $\ST$ by Cauchy surfaces is compatible with the action of the global time translation isometry group of $\ST$.

But this is not the only role of primitive causality. It makes the (covariant) dynamics on the observables \textit{deterministic}. Since a state on a single Cauchy surface algebra $\OA_\CS$, i.e. a prescription of `initial (expectation) values', fixes already the state on the whole quasilocal algebra $\OA$ the expectation values of  the observables at arbitrary times can be given uniquely in terms of the (covariant) time evolution automorphisms of the observable algebra $\OA$ and the `initial' state. 

Although the dynamics $\{\alpha_{t',t}, t,t\in\RE\}$ is deterministic, it is not necessarily \textit{causal}. That is the deterministic dynamics \textit{per se} does not ensure that
\begin{equation}\label{causal_dyn}
(\iota_{t'}^{-1}\circ\iota_t)(\OA(V_t))\subset\OA(V_{t'}),\quad 
  V_{t'}:=\CS_{t'}\cap (J_+(V_t)\cup J_-(V_t)),\ V_t\subset\CS_t; t,t'\in\RE,
\end{equation}
where $V_t:=V\cap\CS_t$ for some $V\in\SDK$ and $J_+(V_t)\cup J_-(V_t)$ is the causal cone of $V_t$, that is the union of its causal future and causal past. The (deterministic) dynamics on the observables meeting the requirement (\ref{causal_dyn}) is called \textit{causal dynamics} on the observables. It means that the `propagation' of local observable algebras under the dynamics respects the causal cone structure of the underlying spacetime. It ensures also that the state on a local algebra $\iota_{t'}(\OA(V_{t`}))$ fixes the state on a local algebra $\iota_t(\OA(V_t))$, if $V_t$ is in the domain of dependence of $V_{t'}$. 

The local and stronger version of primitive causality is 
\begin{enumerate}
\item[7.] \textit{Local primitive causality.} For any globally hyperbolic bounded subspacetime regions $V\in\SDK$, $\OA(V'')=\OA(V)$.\footnote{If $V''\notin\SDK$ this requirement would mean that extending $\SDK$ by the globally hyperbolic bounded subspacetime regions $V'', V\in\SDK$ and defining $\OA(V''):=\OA(V)$ one obtains an extended net of local algebras satisfying isotony, microcausality, and covariance.}
\end{enumerate}
Local primitive causality entails not only primitive causality but also the causality requirement (\ref{causal_dyn}) of the dynamics: given $V_t$ and $V_{t'}$ as in (\ref{causal_dyn}) local primitive causality and isotony (Requirement 1) leads to $\OA\supset\iota_{t'}(\OA(V_{t'}))=\iota_{t'}(\OA(V_{t'}''))\supset\iota_t(\OA(V_t))$. 

We note that if a net satisfies Haag duality for all bounded globally hyperbolic subspacetime regions $V\in\SDK$, then it also satisfies local primitive causality for them:
\begin{equation}
\OA(V)=\OA(V')'\cap\OA=\OA(V''')'\cap\OA=\OA((V'')')'\cap\OA=\OA(V''),\quad V\in\SDK.
\end{equation}
Conversely, requiring Haag duality only for causally complete regions (that is for regions $V\in\SDK$ satisfying $V''=V$) and local primitive causality for all $V\in\SDK$ Haag duality follows for all $V\in\SDK$:
\begin{equation}
\OA(V)=\OA(V'')=\OA((V'')')'\cap\OA=\OA(V''')'\cap\OA=\OA(V')'\cap\OA.
\end{equation}

What can we say in the absence of primitive causality? In case of a generic globally hyperbolic spacetime there is no Cauchy dynamics $\{\alpha_{t,t'}, t,t'\in\RE\}$ on the observables and the Cauchy surface proper subalgebras $\OA_{\CS_t}, t\in\RE$ are not necessarily isomorphic. In case of stationary spacetimes a covariant dynamics $\{\alpha_t,t\in\RE\}\subset\mathrm{Aut}\,\OA$ does exist, however, the isomorphic Cauchy surface subalgebras $\OA_{\CS_t}, t\in\RE$ remain proper subalgebras of $\OA$. Their intersection can be even trivial. Therefore there is no point in speaking about causality of the covariant dynamics, because local subalgebras `propagate' into completely new local subalgebras of $\OA$. Moreover, the covariant dynamics is not deterministic in this case, that is the covariant dynamics and the `initial' state $\phi_s\colon\OA_{\CS_s}\to\CO$ does not fix for $t\not=s$ the expectation values of the isomorphic but not identical proper subalgebras $\OA_{\CS_t}$ of $\OA$. 
Hence, either one prescribes the state for the whole quasilocal algebra $\OA$ or an extension of the initial state $\phi_s$ from $\OA_{\CS_s}$ to $\OA$ is needed. In the first case no property forbids a generic state to reveal acausal properties. However, in the latter case properly chosen causal restrictions on the state extension procedure may lead to a subclass of states obeying causal properties. Unfortunately, we do not know how to do such a state extension in case of a LQT. However, in LCTs, where \textit{conditional probabilities} of local observables have a meaning and they provide local extensions of a state, a state extension procedure can be interpreted in terms of a \textit{stochastic dynamics}, where the mentioned conditional probabilities are given by the \textit{transition probabilities} of the underlying stochastic process. To this end there is no need for a covariant dynamics on the classical observables either. Of course, this would ensure the isomorphisms of the image $\sigma$-algebras of the random
variables on the different Cauchy surfaces in the underlying stochastic process, however a stochastic process can be defined without such isomorphisms.

Clearly, any requirement on the state extension procedure in LCTs coming from causality becomes a restriction on the stochastic dynamics. Stochastic dynamics is an existing and well-established research field in general (that is not necessarily local) \textit{classical} theories (Karlin and Taylor, 1975). In case of LQTs we do not know how to do a causal state extension process, therefore we cannot know about its possible (stochastic) interpretation either.\footnote{There exist quantum \textit{mechanical} models with prescribed \textit{stochastic} and not unitary time evolution (Károlyházy, 1966; Ghirardi, Rimini and Weber, 1986; Diósi 1989). However, they are not \textit{local} theories in our sense, and `primitive causality' holds there in the sense that the `observable algebra' is the same for all time slices.} Hence, all of our attempts and examples for establishing a \textit{causal} stochastic dynamics interpretation of the state extension in the absence of primitive causality are within the frame of LCTs.
In the rest of this section we define what is meant by causal stochastic dynamics in LCTs. We use the language of random variables and stochastic dynamics here because certain notions will have a meaning in terms of local $\sigma$-algebras or local abelian von Neumann algebras only if the stochastic process obeys certain local causality requirements. 

Let $\{X_t,t\in\RE\}$ be random variables indexed by the global time parameter of a foliation $\{\CS_t,t\in\RE\}$ of $\ST$ by Cauchy surfaces. The image $\sigma$-algebra $(\CONF_t,\Sigma_t)$ of the measurable map $X_t$, i.e. the random variable is thought to be the (sub-)$\sigma$-algebra of the power set of classical field configurations $\CONF_t$ on the Cauchy surface $\CS_t$. In case of a covariant dynamics the image $\sigma$-algebras $(\CONF_t,\Sigma_t)$ of $X_t$ are isomorphic for all $t\in\RE$. 
The map $X_t$ is given only for the initial time, $X_s\colon(\Omega,\sigma,p)\to(\CONF_s,\Sigma_s)$, that is only the probabilities of the elements $C\in\Sigma_s$ are known, they are given by the probabilities of the inverse images $p(X_s^{-1}(C))$. It is the stochastic dynamics which provides the explicit maps $X_t$, that is the probabilities of sets of configurations,  for $t\not=s$. The stochastic dynamics is given in terms of transition probabilities
\begin{equation}\label{tr_prob}
\mathrm{Pr}\{ X_t\in C_{V(t)}\vert X_{t_i}=x_i\in\CONF_{t_i},i=1,\dots, n\},
\quad t_1< t_2<\dots <t_n< t,
\end{equation}
where $C_{V(t)}\in\Sigma_t$ \textit{is local}, namely, it is a cylindrical set of field configurations on the \textit{bounded} piece $V(t):=V\cap\CS_t,V\in\SDK$ of a Cauchy surface $\CS_t$. Observe that, in face of the denotation, the transition probabilities are \textit{not} necessarily conditional probabilities on local $\sigma$-algebras since the set $\{x_i\}$ containing a single field configuration on the whole Cauchy surface $\CS_{t_i}$ is not local, even it is not necessarily in $\Sigma_{t_i}$. The subsequent requirements are introduced just to make (\ref{tr_prob}) to be a conditional probability on local $\sigma$-algebras, which allows the stochastic dynamics to be interpreted as a state extension procedure from the initial Cauchy surface algebra $\OA_{\CS_s}$ to the whole quasilocal algebra $\OA$.

The stochastic dynamics will be called \textit{causal} if the transition probability of a conditioned local configuration set depends only on configurations on its causal past: 
\begin{equation}\label{causal_proc}
\mathrm{Pr}\{ X_t\in C_{V(t)}\vert X_{t_i}=x_i,i=1,\dots, n\}= \mathrm{Pr}\{ X_t\in C_{V(t)}\vert (X_{t_i}=x_i)_{\vert J_-(V(t))},i=1,\dots, n\},
\end{equation}
where $J_-(V(t))$ is the causal past of $V(t)$ and the subscript $\vert J_-(V(t))$ means that the prescription of the values of the random variables $X_{t_i}$ is restricted to the Cauchy surface piece $\CS_{t_i}\cap J_-(V(t))$. Note, that the right hand side of (\ref{causal_proc}) is the same for any choice of configurations from the cylindrical sets $C_{J_-(V(t))\cap\CS_{t_i}}(x_i) \in\Sigma_{t_i}, i=1,\dots, n$ obtained by the images of the mapping $Z_V$ in (\ref{cylind_set}) of the single configuration $\{ x_i\}$ with $V=C_{J_-(V(t))\cap\CS_{t_i}}, i=1,\dots, n$. Therefore in case of a causal process it is meaningful to consider the transition probabilities as depending only on the intersection of the cylindrical sets $C_{J_-(V(t))\cap\CS_{t_i}}(x_i)\in\Sigma_{t_i}$ of the configurations $x_i\in\CONF_{t_i}, i=1,\dots, n$. 

In the presence of a covariant dynamics on the observables we assume that (\ref{tr_prob}) are \textit{stationary transition probabilities}, i.e. they depend only on the differences $t_1-t,\dots, t_n-t$. We will examine only \textit {Markov processes}, where only the `closest' conditioning counts, that is
\begin{equation}\label{markov_proc}
\mathrm{Pr}\{ X_t\in C_{V(t)}\vert X_{t_i}=x_i,i=1,\dots, n\} = \mathrm{Pr}\{ X_t\in C_{V(t)}\vert X_{t_n}=x_n\}
\end{equation}
holds whenever $t_1< t_2<\dots <t_n< t$. In case of a causal Markov process the transition probabilities (\ref{tr_prob}) are called \textit{independent with respect to spacelike separation} if the following property holds: Let $V(t)$ be a finite union of disjoint regions $V_k(t):=V_k\cap\CS_t, V_k\in\SDK, k=1,\dots, r$ on the Cauchy surface $\CS_t$ such that their `causal shadows' $J_-(V_k(t))\cap\CS_s$ are also disjoint regions in the Cauchy surface $\CS_s$, i.e. they are also \textit{spacelike separated}. Then the transition probability becomes a product of transition probabilities 
\begin{equation}\label{ind_cm_proc}
\mathrm{Pr}\{ X_t\in C_{V(t)}\vert X_s=x_s\} = \prod_{k=1}^r\mathrm{Pr}\{ X_t\in C_{V_k(t)}\vert X_s=x_s\}
\end{equation}
corresponding to the spacelike separable regions.

The important role of causality property (\ref{causal_proc}) is that the transition probabilities (\ref{markov_proc}) of the Markov process depend only on the equivalence class, the cylindrical set, $C_{J_-(V(t))\cap\CS_{t_n}}(x_{t_n})\in\Sigma_{t_n}$ of the configuration $x_{t_n}\in\CONF_{t_n}$ thus they can be interpreted as conditional probabilities. Hence, they can serve as a state (probability measure) extension procedure of the initial state\footnote{The random variable $X_s$ is a measurable map from the probability space $(\Omega,\Sigma,p)$ into the $\sigma$-algebra $(\CONF_s,\Sigma_s)$.} $\phi_s:=p\circ X_s^{-1}$ on $\Sigma_s$ to the state $\phi$ on the $\sigma$-algebra generated by $\Sigma_t, t\geq s$:
\begin{equation}\label{stoch_state_ext}
\phi(C_{V(t)}\cap C_{J_-(V(t))\cap\CS_s}(x_s)):= 
\mathrm{Pr}\{ X_t\in C_{V(t)}\vert X_s=x_s\}\,
\phi_s(C_{J_-(V(t))\cap\CS_s}(x_s)). 
\end{equation}
Therefore a fortiori the equality (\ref{stoch_state_ext}) implies that the the transition probability is equal to the conditional probability
\begin{eqnarray}\label{transprob_condprob}
\mathrm{Pr}\{ X_t\in C_{V(t)}\vert X_s=x_s\}&=&
\frac{\phi(C_{V(t)}\cap C_{J_-(V(t))\cap\CS_s}(x_s))} 
{\phi_s(C_{J_-(V(t))\cap\CS_s}(x_s))}
=\frac{\phi(C_{V(t)}\cap C_{J_-(V(t))\cap\CS_s}(x_s))} 
{\phi(C_{J_-(V(t))\cap\CS_s}(x_s))}\nonumber\\
&=:&\phi(C_{V(t)}\cap C_{J_-(V(t))\cap\CS_s}(x_s)\vert 
  C_{J_-(V(t))\cap\CS_s}(x_s)), 
\end{eqnarray}
which is possible only in case of a causal process.

We do not know whether Bell's local causality holds in an arbitrary LCT equipped with a state obtained by a causal Markov process with stationary transition probabilities obeying independence with respect to spacelike separation. Nevertheless, this implication holds in LCTs with locally finite dimensional Neumann algebras, which we prove in Section 6.

\section{Further relativistic causality principles} 

Before turning to Bell's local causality principle and its relation to (local) primitive causality in this section we briefly review some other relativistic causality principles present in the literature and their relations to (local) primitive causality. These principles are formulated in a quasilocal algebra $\OA_\HS$ generated by an isotone (Requirement 1) net $\{\vNA(V),V\in\SDK\}$ of local von Neumann algebras.

Let $\left\{ A_k \right\}_{k\in K}\subset\vNA(V_A)$  be a decomposition of the unit, that is a set of mutually orthogonal projections in the local von Neumann algebra $\vNA(V_A)$ such that $\sum_k A_k = \UN$. The corresponding \textit{non-selective} projective measurement is defined as a map  $\OP_{\{A_k\}}\colon\OA_\HS\to\OA_\HS$
\begin{equation}\label{nonselect}
\OP_{\{A_k\}}(X):=\sum_{k\in K} A_k X A_k,\quad X\in\OA_\HS.
\end{equation}
Being a unit preserving completely positive map (even a conditional expectation) $\OP_{\{A_k\}}$ maps states to states via 
\begin{equation}\label{nonselectphi}
\phi \mapsto \phi_{\{A_k\}} := \phi \circ \OP_{\{A_k\}}.
\end{equation}
The following causality principle requires that projections (quantum events) located in spatially separated regions should be insensitive of such a change of states:
\begin{enumerate}
\item[8.] \textit{No-signaling} (also called as \textit{parameter independence}). (Shimony, 1986) Let $V_A, V_B\in\SDK$ be spacelike separated. For any decomposition of the unit $\left\{ A_k \right\}_{k\in K}\subset\vNA(V_A)$ and projection $B\in \vNA(V_B)$, and for any locally faithful and normal state $\phi\colon\OA_\HS\to\CO$, we have
\begin{equation}\label{nosign}
\phi_{\{A_k\}}(B) = \phi(B)
\end{equation}
\end{enumerate}
No-signaling follows from microcausality (Requirement 2). Schlieder (1969) showed that the converse also holds: if no-signaling holds for a decomposition of the unit $\left\{ A_k \right\}_{k\in K}$ and a projection $B$ for all normal states of a von Neumann algebra, then $[A_k,B]=0$ for all $k\in K$. Being equivalent to microcausality no-signaling trivially fulfils in LCTs. Although it is formulated as a requirement for states, it gives a restriction for the structure of the local algebras.

Instead of non-selective projective measurements (\ref{nonselect}) one can also consider \textit{selective projective measurements} using a single local projection $A\in\vNA(A)$:
\begin{equation}\label{select}
\OP_A(X):= AXA,\quad X\in\OA_\HS,
\end{equation}
which defines a completely positive but not unit preserving map $\OP_A\colon\OA_\HS\to\OA_\HS$. The generated state transition 
\begin{equation}\label{selectphi}
\phi \mapsto \phi_{A} :=  \frac{\phi \circ \OP_{A}}{\phi(A)} 
=\frac{\phi \circ \OP_A}{(\phi\circ\OP_A)(\UN)}
\end{equation}
sometimes called \textit{Lüders projection} (Lüders 1950), provides another causality requirement:
\begin{enumerate}
\item[9.] \textit{Outcome independence.} (Shimony, 1986) For any projections $A\in\vNA(V_A)$ and $B\in \vNA(V_B)$ such that $V_A, V_B\in\SDK$ are spacelike separated regions, and for any locally faithful and normal state $\phi$, we have
\begin{equation}\label{outcome}
\phi_{A}(B) = \phi(B)
\end{equation}
\end{enumerate}
In case of microcausality (Requirement 2), outcome independence implies that $\phi(AB) = \phi(A)\phi(B)$, that is $\phi$ becomes a product state by restricting it to the subalgebra generated by $\vNA(V_A)$ and $\vNA(V_B)$. Hence, it is a too strong assumption, which is violated in LQTs, for example, by any entangled state. Of course, it is violated also in case of superluminal correlations.

In general, (completely) positive maps $\OP\colon\OA\to\OA$ on a $C^*$-algebra $\OA$ with the property $0<\OP(\UN)\leqslant\UN$ can be considered as generalized measurements or \textit{operations}. They are called \textit{inner} if $\OP$ has the form $\OP:=\sum_i {\mathrm Ad}\, K_i$ with $K_i\in\OA$. If the $K_i$-s are mutually orthogonal projections one speaks about \textit{projective} (inner) operations. Operations with  $\OP(\UN)= \UN$ and $\OP(\UN)<\UN$ are called\textit{ non-selective} and \textit{selective} operations, respectively. If $\OA$ is a von Neumann algebra one usually requires $T$ to be normal. If $\OA=\BO(\HS)$ this means that $T$ is $\sigma$-weakly continuous. See e.g. (Werner, 1987) and references therein.

A net satisfying local primitive causality (Requirement 7) also satisfies: 
\begin{enumerate}
\item[10.] \textit{Local determinism.} (Earman and Valente, 2014) For any two states $\phi$ and $\phi'$ and for any globally hyperbolic spacetime region $V\in\SDK$, if $\phi|_{\OA(V)}=\phi'|_{\OA(V)}$ then $\phi|_{\OA(V'')}=\phi'|_{\OA(V'')}$
\end{enumerate}
and consequently it also satisfies 
\begin{enumerate}
\item[11.] \textit{Stochastic Einstein locality.} Let $V_A, V_C\in\SDK$ such that $V_C \subset J_{-}(V_A)$ and $V_A \subset V''_C$. 
If $\phi|_{\OA(V_C)}=\phi'|_{\OA(V_C)}$ holds for any two states $\phi$ and $\phi'$ on $\OA$ then $\phi(A) = \phi'(A)$ for any projection $A \in \OA(V_A)$.
\end{enumerate}

Microcausality alone does \textit{not} entail local primitive causality. Since microcausality is equivalent to no-signaling and local primitive causality represents no-superluminal propagation (Earman and Valente, 2014), therefore it is an interesting question whether there exist nets which \textit{satisfy} local primitive causality but \textit{violate} microcausality. Usually the translation covariant field algebra extension of the observables $\FA\supset\OA$, in which the localized and transportable endomorphisms--- the Doplicher--Haag--Roberts morphisms---of the observables can be implemented, serve such examples: Although local field algebras are defined to be relatively local to observables 
\begin{eqnarray}
\FA(V):=\OA(V')'\cap\FA,\quad V\in\SDK, 
\end{eqnarray}
local field algebras corresponding to spacelike separated regions do not commute in general, hence microcausality fails. (For example, in the field algebra of the local quantum Ising model there are field operators with spacelike separated supports that anticommute.) However, local primitive causality does hold in the net of field algebras, because $V'=V'''$ and hence
\begin{eqnarray} 
\FA(V):=\OA(V')'\cap\FA=\OA(V''')'\cap\FA=\OA((V'')')'\cap\FA=:\FA(V''),
\quad V\in\SDK.
\end{eqnarray}
Thus, for such a net of local (field) algebras no-signaling is violated whereas no-superluminal propagation holds.

In the following we will work within the framework of a LPT. When speaking about deterministic dynamics, we will also assume Requirements 6-7.

\section{Bell's notion of local causality}

Local causality has been one of the central notions in Bell's writings on the foundations of quantum mechanics. Still, interestingly the notion of local causality gets an explicit formulation only in few of  his papers; to our knowledge only in (Bell, 1975/2004, p. 54), (Bell, 1986/2004, p. 200), and (Bell, 1990/2004, p. 239-240). In this latter posthumously published paper, ``La nouvelle cuisine'', local causality is formulated as follows:\footnote{For the sake of uniformity throughout the paper we slightly changed Bell's denotation and figures.}
\begin{quote}
``A theory will be said to be locally causal if the probabilities attached to values of local beables in a space-time region $V_A$ are unaltered by specification of values of local beables in a space-like separated region $V_B$, when what happens in the backward light cone of $V_A$ is already sufficiently specified, for example by a full specification of local beables in a space-time region $V_C$. ''  (Bell, 1990/2004, p. 239-240)
\end{quote}
\begin{figure}[ht]
\centerline{\resizebox{10cm}{!}{\includegraphics{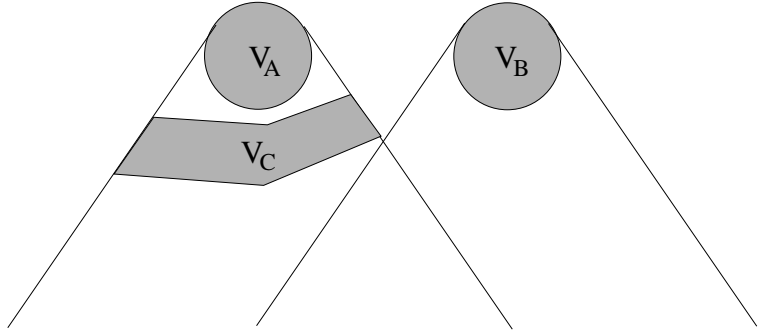}}}
\caption{Full specification of what happens in $V_C$ makes events in $V_B$ irrelevant for predictions about $V_A$ in a locally causal theory.}
\label{loc2v}
\end{figure}
The figure Bell is attaching to this formulation is reproduced in Fig. \ref{loc2v} with the original caption. Bell elaborates on his formulation as follows:
\begin{quote}
``It is important that region $V_C$ completely shields off from $V_A$ the overlap of the backward light cones of $V_A$ and $V_B$. And it is important that events in $V_C$ be specified completely. Otherwise the traces in region $V_B$ of causes of events in $V_A$ could well supplement whatever else was being used for calculating probabilities about $V_A$. The hypothesis is that any such information about $V_B$ becomes redundant when $V_C$ is specified completely.''  (Bell, 1990/2004, p. 240)
\end{quote}
The notions featuring in Bell's formulation has been target of intensive discussion in philosophy of science (see Norsen 2009, 2011). Here we would like to concentrate only on three terms, namely \textit{local beables}, \textit{complete specification} and \textit{shielding-off}.
\vspace{0.1in}

\noindent
\textit{Local beables}. The notion ``beable'' is Bell's neologism and is contrasted to the term ``observable'' used in quantum theory. ``The \textit{be}ables of the theory are those entities in it which are, at least tentatively, to be taken seriously, as corresponding to something real'' (Bell, 1990/2004, p. 234). The clarification of what the ``beables'' of a theory are, is indispensable in order to define local causality since 
``there \textit{are} things which do go faster than light. British sovereignty is the classical example. When the Queen dies in London 
(long may it be delayed) the Prince of Wales, lecturing on modern architecture in Australia, becomes instantaneously King'' (p. 236).

Beables are to be local: ``\textit{Local} beables are those which are definitely associated with particular space-time regions. The electric and magnetic fields of classical electromagnetism, ${\bf E}(t, x)$ and ${\bf B}(t, x)$ are again examples.'' (p. 234). 
\vspace{0.1in}

\noindent
\textit{Complete specification}. Local beables are to ``specify completely'' region $V_C$ in order to block causal influences arriving at $V_A$ from the common past of $V_A$ and $V_B$. (For the question of \textit{complete} \textit{vs}. \textit{sufficient} specification see  (Norsen, 2011; Seevinck and Uffink 2011; Hofer-Szabó 2015a).)
\vspace{0.1in}

\begin{figure}[ht]
\centerline{\resizebox{10cm}{!}{\includegraphics{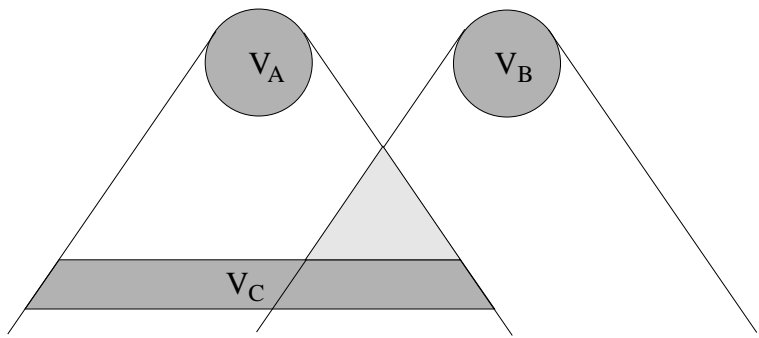}}}
\caption{A \textit{not} completely shielding-off region $V_C$.}
\label{loc2_4}
\end{figure}

\noindent
\textit{Shielding-off}. ``It is important that region $V_C$ completely shields off from $V_A$ the overlap of the backward light cones of $V_A$ and $V_B$.'' Why is that so? Why local causality is not required for such regions $V_C$ as depicted in Fig. \ref{loc2_4}, for example?
The reason for that is the following. If $V_C$ is localized as in Fig. \ref{loc2_4}, then the spacetime region \textit{above} $V_C$ in the common past of the correlating events may contain stochastic events (with determined probabilities by the complete specification on the region $V_C $) which can establish a correlation between $A$ and $B$ in a classical stochastic theory. The ``shielding-off'' condition is required just to exclude this case.

\begin{figure}[ht]
\centerline{\resizebox{10cm}{!}{\includegraphics{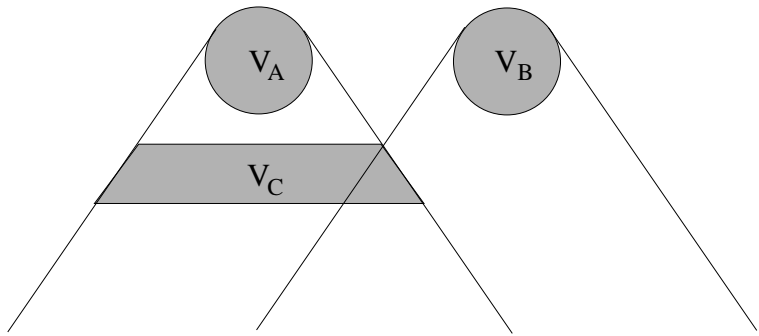}}}
\caption{An intersecting and completely shielding-off region $V_C$.}
\label{loc22}
\end{figure}

But if this is the reason, then why not to allow also for regions $V_C$ as depicted in Fig. \ref{loc22}?
Allowing for shielding-off regions which \textit{intersect} with the common past is indeed a possible interpretation of Bell's term ``shielding-off''. We will return to this point below. (For the relation between the localization of the region $V_C$ and the Causal Markov Condition see  (Hofer-Szabó 2015b).)
\vspace{0.1in}

\noindent
How to translate Bell's three above terms into the framework of LPT? Let us see them again in turn.
\vspace{0.1in}

\noindent
\textit{Local beables}. In a classical field theory beables are characterized by sets of field configurations. In our local algebraic framework local equivalence classes of field configurations, namely, configurations having the same field values on a given spacetime region, generate local $\sigma$-algebras, as explained in Section 2. The elements of local $\sigma$-algebras capture all the beables of the theory, moreover they also provide a localization for them. Translating $\sigma$-algebras into abelian von Neumann algebras one can use a common language for classical and quantum theories: ``local beables'' in a region $V\in\SDK$ are \textit{elements of the local von Neumann algebra} $\vNA(V)$, which is abelian for a classical and non-abelian for a quantum theory.
\vspace{0.1in}

\noindent
\textit{Complete specification}. Complete specification of field configurations in a given spacetime region means that one specifies the field values to a prescribed value in the given spacetime region, that is one specifies the corresponding local equivalence class (a cylindrical set) of a single configuration. In probabilistic language complete specification is translated to a probability measure having support on this local equivalence class of the single specified configuration. More precisely, complete specification is such a change of the probability measure on the whole $\sigma$-algebra that the resulted probability measure restricted to the  local $\sigma$-algebra in question will have support on the local equivalence class of the single specified configuration. In the abelian von Neumann language this corresponds to a change of the original state that results in a pure state on the local von Neumann algebra in question with value 1 on the projection corresponding to the local equivalence class of the single specified configuration. 
However, we would like also this change of states to be as local as possible. Therefore we translate a ``complete specification of beables in a region $V\in\SDK$'' as a change of state
\begin{equation}\label{change}
\phi(X) \mapsto \phi_{\OP} (X) :=  \frac{\phi\circ\OP}{(\phi\circ\OP)(\UN)} 
\end{equation}
by a completely positive map $\OP$ on the quasilocal observables obeying the following properties:
\begin{enumerate}
\item[]{$\textbf P_1:$} the restriction of $\phi_{\OP}$ to the local algebra $\vNA(V)$ is pure,  
\item[]{$\textbf P_2:$} $B\OP(\UN)=\OP(B)=\OP(\UN)B$ hold for local observables $B$ supported in $V'$.
\end{enumerate}
Concerning property $\textbf P_1$ we note that von Neumann algebras in $\BO(\HS)$ which have a separating vector in $\HS$, irrespectively of being abelian or non-abelian algebras, do not possess a pure normal state (Clifton and Halvorson, 2001). This is the case, for example, in AQFTs with type III local von Neumann algebras. Thus starting from a (locally) normal state $\phi$ on them a normal operation $\OP$ leads to a (locally) normal state $\phi_\OP$ which cannot be pure. There are two ways to circumvent this problem (none of them being fully satisfactory): 1. One can use a non-normal operation to get a pure state for the local von Neumann algebra. In this case, however, one jumps into a different quasi-equivalence class of representations of observables which we just wanted to avoid by considering only (locally) normal states for the local von Neumann algebras. 2. In case of type III (hence non-abelian) local von Neumann algebras one can also assume the split property (see e.g. (Werner, 1987) and references therein) and use the (atomic) type I intermediate von Neumann algebra to provide a pure state, hence a `full specification', for a somewhat larger local observable algebra supported in a somewhat larger local region.\footnote{The authors thank to Yuichiro Kitajima for drawing their attention to these points.}

Concerning property $\textbf P_2$ we note that \textit{weakly localized} operations in $V$ (Werner, 1987) obey property $P_2$ for all elements  $B\in\vNA(V)'\supseteq \OA_\HS(V')$ by definition. Moreover, if $\OP$ is normal and $\OA_\HS=\BO(\HS)$ then every weakly localized operation $\OP$ with respect to $V\in\SDK$ is inner in $\vNA(V)$, that is $\OP=\sum_i {\mathrm Ad}\, K_i $ with $K_i\in\vNA(V)$.

In a general LPT, we do not know how to characterize the operations that result in a state obeying properties $P_1$ and $P_2$, but in case of atomic (type I) local von Neumann algebras it is almost trivial: one has to do a selective projective measurement defined in (\ref{select}) by an atom (a minimal projection) $C$ in the local algebra $\vNA(V)$ which induces the change of states $\phi\mapsto \phi_C$ defined in (\ref{selectphi}).
\vspace{0.1in}

\noindent
\textit{Shielding-off}. Finally, a shielding-off region in a LQT (see Fig. \ref{loc2v}) can be defined as $V_C\in\SDK$ satisfying the following three localization requirements:
\begin{enumerate}
\item[]{$\textbf L_1:$} $V_C \subset   J_-(V_A)$,  
\item[]{$\textbf L_2:$} $V_A \subset V''_C$,
\item[]{$\textbf L_3^Q:$} $V_C\subset V_B'$. 
\end{enumerate}
In a LCT a shielding-off region intersecting with the common past (see Fig. \ref{loc22}) is allowed, and requirement $L_3^Q$ can be replaced by the weaker requirement:
\begin{enumerate}
\item[]{$\textbf L_3^C:$} $J_-(V_C)\supset J_-(V_A) \cap J_-(V_B)$.
\end{enumerate}
In case of a Cauchy algebra of an infinitely thin Cauchy surface, requirement $L_3^C$ coincides with requirement $L_3^Q$.
\vspace{0.1in}

\noindent
Given the above translations of the terms ``local beables'', ``complete specification'' and ``shielding-off,'' now we are in the position to formulate Bell's notion of local causality in the framework of LPTs:
\begin{D}\label{BellLC} Let an LPT represented by a net $\{\vNA(V),V\in\SDK\}$ of von Neumann algebras. Let $A \in\mathcal \vNA(V_A)$ and $B\in\mathcal \vNA(V_B)$ be a pair of projections supported in spacelike separated regions $V_A, V_B\in\SDK$. Let $\phi$ be a locally normal and locally faithful state on the quasilocal observables establishing a correlation $\phi(AB)\neq \phi(A)\phi(B)$ between $A$ and $B$. Let $\OP$ be an operation on the quasilocal observables obeying properties $P_1$ and $P_2$. Finally, let $V_C\in\SDK$ be a spacetime region defined by requirements $L_1$, $L_2$ and $L^Q_3/L^C_3$. The LPT is called \textit{(Bell) locally causal} if for any such quintuple $(A,B,\phi,\OP,V_C)$ the following screening property holds:
\begin{equation}\label{BLC_screening}
\phi_{\OP}(AB)=\phi_{\OP}(A)\phi_{\OP}(B).
\end{equation}
\end{D}
\noindent
\textbf{Remarks:}
\begin{enumerate}
\item If the local algebras of the net are atomic,\footnote{Which is typically not the case in a general AQFT.} the states $\phi_{\OP}$ in Definition \ref{BellLC} can be replaced by the state $\phi_C$ given by (\ref{select}--\ref{selectphi}), where $C\in\OA(V_C)$ is an arbitrary atomic event, i.e. a minimal projection. This converts (\ref{BLC_screening}) into the screening-off property:
\begin{eqnarray}\label{BLC_screening_sandwich}
\frac{\phi(CABC)}{\phi(C)} = \frac{\phi(CAC)}{\phi(C)}\frac{\phi(CBC)}{\phi(C)}.
\end{eqnarray}
In LCTs this can be written into the well-known conditional form
\begin{eqnarray}
p(AB\vert C) = p(A\vert C)p(B\vert C),\label{BLC''}
\end{eqnarray}
or into the equivalent asymmetric form
\begin{eqnarray}
p(A\vert BC) = p(A\vert C)\label{BLC'''}
\end{eqnarray}
sometimes used in the literature (for example in (Bell, 1975/2004, p. 54)).
\item Here we would like to briefly comment on a definition of local causality recently given by Joe Henson (2013b). Henson's definition differs from ours in three respects: First, Henson formulates local causality in terms of \textit{$\sigma$-algebras}. Using the recipe given in Section 2 to convert $\sigma$-algebras into abelian von Neumann algebras this difference can be easily dissolved. Second, Henson definition applies only to \textit{atomic} $\sigma$-algebras: his screening-off condition is equivalent to  (\ref{BLC_screening_sandwich}). Our more general screening condition  (\ref{BLC_screening}) applies both to noncommutative and to nonatomic local algebras. Third, in Henson's definition the screener-off region $V_C$ is not localized according to requirements $L_1$, $L_2$ and $L^Q_3/L^C_3$. It is an \textit{unbounded} region, a ``suitable past'' of $V_A$ and $V_B$.\footnote{Where the term ``suitable past''  ``has been left open deliberately.'' ''It could be \dots the 'mutual past' \dots 
the 'joint past' or the past of one of the regions but not the other.'' (Henson, 2013b, p. 1015) For an argument \textit{for}, \textit{against} and again \textit{for} not specifying the screener-off region see (Henson, 2005), (Rédei and San Pedro, 2012) and (Henson, 2013a), respectively.} 
In our opinion, Henson follows here Bell's \textit{first} formulation of local causality given in (Bell, 1975/2004, p. 54), where the screener-off regions are identified with the \textit{complete, unbounded causal past} of the correlating events. Our definition, on the other hand, is based on Bell's \textit{last}, operationally more desirable definition provided in (Bell, 1990/2004, p. 239-240), where the screener-off regions are only \textit{bounded Cauchy segments} of the unbounded past regions.\footnote{Cf. also (Bell, 1986/2004, p. 200): ``The notion of local causality presented in this reference [namely in (Bell, 1975/2004)] involves complete specification of the beables in an infinite space-time region. The following conception is more attractive in this respect.” And then comes the new definition based on bounded regions.} (For a comparison of Bell's different versions of local causality see (Hofer-Szabó 2015b).)

In his paper Henson shows that the lack of separability (\textit{additivity}, in our language, see Section 2) does not block the derivation of the Bell inequalities. As we will see, this result is in complete agreement with ours: additivity is not required in our paper, hence it plays no role in the derivation of the Bell inequalities in LCTs.
\end{enumerate}
Coming back to Definition \ref{BellLC} of local causality, the main question is that when a LPT is locally causal? We answer this question by the following
\begin{Prop}\label{BLC_LPC} Let the local von Neumann algebras of a LPT be atomic. Then Bell's local causality holds if the LPT obeys local primitive causality.
\end{Prop}
\noindent\textit{Proof.} If $A$ is a projection and $C$ is a minimal projection in an atomic von Neumann algebra then $CAC=r(C,A)C$ with $r(C,A)\in\{0,1\}$ in case of abelian and $r(C,A)\in[0,1]\subset\RE$ in case of non-abelian algebras. Hence, using notations of Definition \ref{BellLC}, $A$ is a projection in the atomic von Neumann algebra $\vNA(V_C)$ due to local primitive causality. Thus if $C\in\vNA(V_C)$ is a minimal projection then 
\begin{eqnarray}\label{BLC_LPT_screening}
\phi_C(AB)&:=&\frac{\phi(CABC)}{\phi(C)} =\frac{\phi(CACB)}{\phi(C)} = r(C,A)\frac{\phi(CB)}{\phi(C)}= 
\frac{\phi( CAC)}{\phi(C)}\frac{\phi(CBC)}{\phi(C)}\nonumber\\
&=:&\phi_C(A)\phi_C(B).
\end{eqnarray}
Here we used that $CB=BC$ due to commutativity in case of a LCT and due to the spacelike separation of $V_B$ and $V_C$ (ensured by requirement $L_3^Q$) and microcausality in a LQT.\qed
\vspace{0.1in}

\noindent
In the light of this proposition the reader may ask how a local \textit{quantum} theory can be locally causal if local causality implies various Bell inequalities which are known to be violated for certain set of quantum correlations. We come back to this point in Section 7.

In case of LPTs with local primitive causality but with \textit{non-atomic} von Neumann algebras we do not know how to characterize the local manipulation on the state described in Definition \ref{BellLC}, therefore a similar proof cannot be applied. In case of LPTs \textit{without local primitive causality} the dynamics is not deterministic, hence an initial state on a Cauchy surface algebra does not determine the state on the whole quasilocal algebra $\OA$. States can be forced by a properly chosen state extension procedure to show suitable causality properties. We will not investigate such state extensions in LQTs but only in LCTs where the extension procedure can be interpreted as a causal stochastic dynamics on the states. LCTs equipped by such states will be called \textit{stochastic LCTs} for short. In the next section we consider their relation to Bell's local causality and a simple prototype of them, the causal stochastic Ising model will be constructed.

\section{Bell's local causality in stochastic LCTs}

We start the section by a
\begin{Prop}\label{BLC_stoch_LPC} 
Let the stochastic dynamics in a LCT (without primitive causality) be given by a stationary causal Markov process with transition probabilities independent with respect to spacelike separation defined in Section 3. Let the local von Neumann algebras of the LCT be finite dimensional. Then Bell's local causality holds for any region $V_C$ allowed by Definition \ref{BellLC}.
\end{Prop}
\noindent\textit{Proof.}  Let $\phi\colon\OA(s,s')\to\CO$ be the state on a time interval quasilocal observable algebra extended from a state $\phi_s$ on the Cauchy surface algebra $\OA_s$ by the stochastic process. Let $A,B\in\OA(s,s')$ and $C=C_t\tilde C\in\OA(s,s')$ be given as in Definition \ref{BellLC} such that $C_t$ is a minimal projection in a Cauchy surface algebra $\OA_t$ obeying $V_{C_t}''\supset V_A$, and $J_-(V_{C_t})\supset V_{\tilde C}$. Let $\{D^k_t\}\subset \OA(\CS_t\cap J_-(V_B))$ be the (finite) partition of unit into minimal projections. Then using (\ref{transprob_condprob}), Markov property, and the independence of transition probabilities with respect to spacelike separation one obtains 
\begin{eqnarray}\label{BLC_stoch_screening}
\phi_C(AB)&:=&\frac{\phi(ABC)}{\phi(C)} 
 =\sum_k\frac{\phi(ABCD^k_t)}{\phi(C)}
 =\sum_k\frac{\phi(ABCD^k_t)}{\phi(CD^k_t)}\frac{\phi(CD^k_t)}{\phi(C)}
 =\sum_k\mathrm{Pr}\{ AB\vert C_t\tilde CD^k_t\}
       \frac{\phi(CD^k_t)}{\phi(C)}\nonumber\\
&=&\sum_k\mathrm{Pr}\{ AB\vert C_tD^k_t\}\frac{\phi(CD^k_t)}{\phi(C)}
= \sum_k\mathrm{Pr}\{ A\vert C_t\}\mathrm{Pr}
\{ B\vert D_t^k\}\frac{\phi(CD^k_t)}{\phi(C)}\nonumber\\
&=&\sum_k\mathrm{Pr}\{ A\vert C\}\mathrm{Pr}\{ B\vert D_t^k\}\frac{\phi(CD_t^k)}{\phi(C)}
= \phi_C(A)\sum_k\mathrm{Pr}\{ B\vert D_t^k\}\phi_C(D_t^k)
= \phi_C(A)\phi_C(B),
\end{eqnarray}
which is the screening condition (\ref{BLC_screening_sandwich}) required by Bell's local causality.\qed
\vspace{0.1in}

\noindent
In the following we present a simple stochastic LCT in $\ST^2$ with finite dimensional local algebras. Since the dynamics is given by a stationary causal Markov process with transition probabilities independent with respect to spacelike separations and since the local algebras on minimal elements of $\SDK$ are two dimensional we call it \textit{causal stochastic Ising model}. We show that due to the prescribed properties of the process the model can be characterized by eight parameters, which are local transition probabilities.

Consider a locally finite covering of the two dimensional Minkowski spacetime $\ST^2$ given by minimal double cones $V^m(t,i)$ of unit diameter with their center in $(t,i)$ for $t,i\in\mathbb{Z}$ or $t,i\in\mathbb{Z}+1/2$. This set of minimal double cones is denoted by $S^m$. A generic double cone $V$ in this discretization is a finite subset of $S^m$ generated by two of its elements: $V\equiv V(t,i;s,j):=V^m(t,i)\vee V^m(s,j)$ is the smallest double cone in $\ST^2$ containing both $V^m(t,i)$ and $V^m(s,j)$. The directed poset of such double cones in $\ST^2$ is denoted by $\SDK^m$.

\begin{figure}[ht]
\centerline{\resizebox{9cm}{!}{\includegraphics{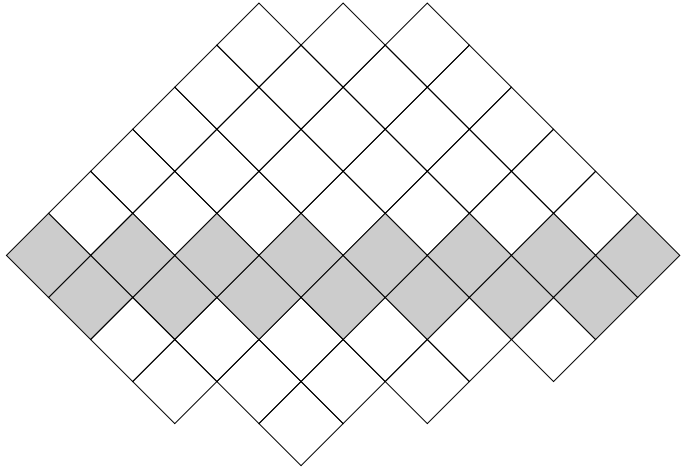}}}
\caption{Locally finite covering of the two dimensional Minkowski spacetime with a `thickened' Cauchy surface.}
\label{LC1}
\end{figure}

Let $S_t^m\subset S^m$ be the subset of minimal double cones with time coordinate $t\in\fel\IN$. Minimal double cones with time coordinates $t$ and $t+\fel$ form a `thickened' Cauchy surface $\CS_t:=S^m_t\cup S^m_{t+\fel}$ in this locally finite covering of $\ST^2$ (see Fig. \ref{LC1}). 
A double cone $V\in\SDK^m$ is sticked to the Cauchy surface $\CS_t$ if it is generated by two minimal double cones in $\CS_t$. The directed poset of double cones sticked to $\CS_t$ is denoted by $\SDK^m_t$, it is contained in $\SDK^m$.  Obviously, $\SDK^m_t$ is left invariant by integer space translations and $\SDK^m$ is left invariant by integer space and time translations.

Let $\IN_2$ be the group with two elements represented by the multiplicative group of the integers $\{1,-1\}$. A $\IN_2$-valued field configuration on this covering of $\ST^2$ is a map $c\colon S^m\to \IN_2$. Using the identification $F=\IN_2$ for field values and $\ST=S^m$ for the underlying spacetime we will follow not only the general construction of a LCT from a classical field theory but also the definition of a causal Markov process with stationary transition probabilities obeying independence with respect to spacelike separation described in Section 3. As a result we arrive at a LCT with a very simple local rule of a stochastic dynamics.

Let $\CONF:=\{ c\colon S^m\to\IN_2\}\equiv \IN_2^{S^m}$ be the set of \textit{field configurations}. The maximal $\sigma$-algebra of classical events one can imagine in this model is $(\CONF,\PR(\CONF))$ given by the power set $\PR(\CONF)$ of the set of field configurations. An isotone net structure $\{(\CONF,\Sigma(V)), V\in\SDK^m\}$ of unital $\sigma$-subalgebras $\Sigma(V)\subset\PR(\CONF)$ labeled by double cones in $\SDK^m$ (or even by elements of $\FMK^m$ being finite subsets of $S^m$) can be given by the `cylindrical subsets' of $\CONF$ corresponding to the image sets of the mappings $Z_V\colon\PR(\CONF)\to\PR(\CONF), V\in\SDK^m$ defined in (\ref{cylind_set})
\begin{equation}\label{Z2_cylind_set}
\PR(\CONF)\ni C\mapsto Z_V(C):=\{ c'\in\CONF \, | \exists c\in C: 
c_{\vert V}= c'_{\vert V}\}\in\Sigma(V):=Z_V(\PR(\CONF)).
\end{equation}
Since $\SDK^m$ is a subset of $\FMK^m$, that is every $V\in\SDK^m$ is a finite subset of $S^m$, the local $\sigma$-algebras are finite. Namely, $\Sigma(V)$ is isomorphic to the power set $\PR(\CONF_V)$ of $\CONF_V$, the set of local equivalence classes of single configurations, where the local, i.e. $V$-dependent equivalence relation introduced in Section 2 is given by the restriction to $V$: $c\sim_V c'$ if $c_{\vert V}=c'_{\vert V}$. Clearly, $\CONF_V$ contains $2^{\vert V\vert}$ elements, where $\vert V\vert$ is the number of minimal double cones in $V$. Note, that the local $V$-equivalence class $C\equiv[c]_V\in\CONF_V$ of a single configuration $c\in\CONF$ is a \textit{minimal} cylindrical subset of $\CONF$ corresponding to $V$ by the map (\ref{Z2_cylind_set}): $[c]_V=Z_V(\{ c\})$, i.e it is an \textit{atom} in $\Sigma(V)$. Hence, the $2^{\vert V\vert}$ dimensional abelian local von Neumann algebra $\vNA(V)$ corresponding to the local $\sigma$-algebra $\Sigma(V)$ is ($\CO$-linearly) spanned by the set of mutually orthogonal
minimal projections $P^C_V,C\in\CONF_V$. 
They correspond to characteristic functions $\chi^C_V\colon\CONF\to\CO$ which are $1$ on the cylindrical subset $C\in\CONF_V$, i.e. on a $V$-equivalence class of a single configuration in $\CONF$, and $0$ otherwise. The local $\sigma$-algebras obey the intersection property
\begin{equation} 
\Sigma(V_1)\cap\Sigma(V_2)=\Sigma(V_1\cap V_2),\quad V_1,V_2\in\FMK^m,
\end{equation} 
especially $\Sigma(V_1)\cap\Sigma(V_2)=\{\emptyset,\CONF\}$ if $V_1\cap V_2=\emptyset$. Of course, the local von Neumann algebras inherit this intersection property. First and last, $\{\vNA(V),V\in\SDK^m\}\subset\{\vNA(V),V\in\FMK^m\}$ is an isotone net of finite dimensional, hence atomic, abelian von Neumann algebras obeying the intersection property not only for spacelike separated regions; that is they define a LCT without local primitive causality.

The quasilocal $C^*$-algebra $\OA$ is given by the inductive limit of the local von Neumann algebras $\vNA(V), V\in\SDK^m$. The unital $C^*$-subalgebras $\OA_t, t\in\fel\IN$ of $\OA$ correspond to the thickened Cauchy surfaces $\CS_t\subset S^m$. Clearly, $\OA$ is an integer time and space translation covariant net, i.e. $\PR_{\SDK^m}=\IN\times\IN$. Moreover, it is also covariant with respect to the `half shift' of coordinates of the minimal double cones: $(t,i)\mapsto (t+\fel,i+\fel)$.\footnote{This transformation corresponds to the Kramers--Wannier duality in the local quantum Ising model.} The covariant dynamics, that is image automorphisms $\alpha(n,0), n\in\IN$ of the mapping $\alpha\colon\PR_{\SDK^m}\to\mathrm{Aut}\,\OA$, maps the Cauchy subalgebra $\OA_t$ onto $\OA_{t+n}$, hence, they are isomorphic subalgebras of $\OA$ for $n\in\IN$. However, their intersection is trivial for $n\not=0$. Therefore primitive causality does not hold in this LCT and the covariant dynamics 
$\{\alpha(n,0),n\in\IN\}\subset\mathrm{Aut}\,\OA$
does not carry any further causal property. Causality will reappear in the state extension procedure from a state $\phi_s\colon\OA_s\to\CO$ on a proper Cauchy subalgebra to a state $\phi$ on the whole quasilocal algebra $\OA$. The extension will be given in terms  of a causal stochastic dynamics described in Section 3. 

\begin{figure}[ht]
\centerline{\resizebox{9cm}{!}{\includegraphics{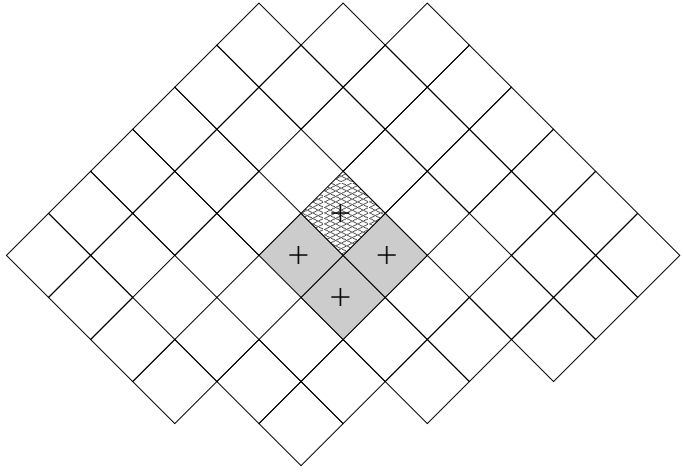}}}
\caption{Three minimal double cones adjacent to $V^m_1$ from below, the configuration on which specifies the transition probabilities.}
\label{LC1_3}
\end{figure}

The set of field configurations on the subset $S^m_t\subset S^m$ of minimal double cones on the time slice $t\in\fel\IN$ is denoted by $\CONF_t$. The image $\sigma$-algebras of the corresponding $\CONF_t$-valued random variables $X_t, t\in\fel\IN$ will be $(\CONF_t,\PR(\CONF_t))$ in this model. As an artifact of the locally finite covering of $\ST^2$ a (thickened) Cauchy surface $\CS_t$ will contain a pair $(X_t,X_{t+\fel})$ of random variables. The discrete stochastic dynamics on the random variables is given by transition probabilities (\ref{tr_prob}) specified to this case as 
\begin{equation}\label{Z2_tr_prob}
\mathrm{Pr}\{ X_t\in C_{V(t)}\vert (X_{t_i},X_{t_i+\fel})=(x_i,x'_i)\in\CONF_{t_i}\times\CONF_{t_i+\fel},i=1,\dots, n\},
\quad t_{i+1}-t_i\geq 1, t-t_n\geq 1,
\end{equation}
where the pairs $(X_{t_i},X_{t_i+\fel})$ correspond to random variables on the Cauchy surface $\CS_{t_i}$ and $V(t)\subset S^m_t$ is a finite set of minimal double cones on the time slice $t$, that is $V(t)\in\FMK_t^m$. The $\IN\times\IN$-covariance of the model allows us to require the transition probabilities to be stationary (time translation invariant) and space translation invariant. Using the notations $Y_t\equiv (X_t,X_{t+\fel})$ and $y\equiv (x,x')$ the Markov condition (\ref{markov_proc}) for the transition probabilities (\ref{Z2_tr_prob}) requires that
\begin{equation}\label{Z2_markov_proc}
\mathrm{Pr}\{ X_t\in C_{V(t)}
  \vert Y_{t_i}=y_i, i=1,\dots, n\} 
= \mathrm{Pr}\{ X_t\in C_{V(t)}\vert Y_{t_n}=y_n\},
\end{equation}
whenever $t_{i+1}-t_i\geq 1$ and $t-t_n\geq 1$ hold. Therefore the `nearest time slice' transition probabilities $\mathrm{Pr}\{ X_1\in C_{V(1)}\vert Y_0=y\}$ completely specify the process if we require invariance of transition probabilities also with respect to the half shift $(t,i)\mapsto (t+\fel,i+\fel)$ of coordinates of the minimal double cones mentioned before. 
The process is required to be causal (\ref{causal_proc}) that is 
\begin{equation}\label{Z2_causal_proc}
\mathrm{Pr}\{ X_1\in C_{V(1)}\vert Y_0=y\}= 
\mathrm{Pr}\{ X_1\in C_{V(1)}\vert (Y_0=y)_{\vert J_-(V(1))}\},
\end{equation}
where $J_-(V(1))$ is the causal past of $V(1)$ and the subscript $\vert J_-(V(1))$ means that the prescription of the values of the random variable $Y_0$ is restricted to the `causal shadow' $\PR_0(V(1))\equiv\CS_0\cap(\CS_0\setminus J_-(V(1)))'$ of $V(1)$ on the Cauchy surface $\CS_0$.\footnote{It is the artifact of the thickened Cauchy surface that the intersection $\CS_0\cap J_-(V(1))$ contains two plus two minimal double cones at the boundary of $J_-(V(1))$ for $V(1)\in\SDK^m_1$. However, the field configuration on the `older' minimal double cones is not needed for a causal transition probability, the relevant double cones are contained in $\CS_0\cap(\CS_0\setminus J_-(V(1)))'$.} We consider only transition probabilities that are independent with respect to spacelike separation, that is they will satisfy (\ref{ind_cm_proc}).\footnote{As an artifact of the thickened Cauchy surface one can choose among different prescriptions which lead to the same condition in case of a `true' 
(infinitely thin) Cauchy surface. Namely, the condition that spacelike separated regions $V_1, V_2\in\SDK$ have spacelike separated shadows $\PR_t(V_1), \PR_t(V_2)$ on the Cauchy surface $\CS_t$ can be formulated as $J_-(V_1)\cap J_-(V_2)\subset J_-(\PR_t(V_1)\cup \PR_t(V_2))$. This prescription is used in (\ref{Z2_ind_cm_proc}).} Since on a single time slice any finite set $V(t)\in\FMK^m_t$ consists of (finite number of) mutually spacelike separated minimal double cones $V^m_t\in V(t)$, we have 
\begin{equation}\label{Z2_ind_cm_proc}
\mathrm{Pr}\{ X_1\in C_{V(1)}\vert 
(Y_0=y)_{\vert J_-(V(1))}\} = \prod_{V^m_1\in V(1)} \mathrm{Pr}\{ X_1\in C_{V^m_1}\vert (Y_0=y)_{\vert J_-(V^m_1)}\}.
\end{equation}
Therefore it is enough to give the transition probabilities $\mathrm{Pr}\{ X_1\in C_{V^m_1}\vert (Y_0=y)_{\vert J_-(V^m_1)}\}$ for a single minimal double cone $V^m_1\in S^m_1$ to specify the process completely. Since the causal shadow $\PR_0(V^m_1)$ of $V^m_1$ on the Cauchy surface $\CS_0$ consists of three minimal double cones (see Fig. \ref{LC1_3}), 
which carry $2^3$ different configurations, we need to specify eight transition probabilities, for example those with $C_{V^m_1}=\{+1\}$. However, the requirement of the existence of unique state extension backward in time restricts not only the possible eight transition probabilities but also the possible `final' states, that is the stochastic process shrinks the possible states that can occur on future Cauchy surface subalgebras: Let $c\in\Sigma_t$ be the set of configurations which is fixed on $V\subset S_t^m$ consisting of two neighboring minimal double cones in $S^m_t$. The configuration sets $c^\pm$ and $c_\pm$ will mean the subset of $c$ where the configurations is fixed to $\pm$ on a third minimal double cone in the future and past domain of dependence region of $V$, respectively. Then the local extension of a state $\phi$ on the Cauchy surface algebra $\OA_{t-\fel}$ to the Cauchy surface algebra $\OA_t$ is given by two dimensional linear mappings:
\begin{equation}
\begin{pmatrix}\phi(c^+)\\ \phi(c^-)\end{pmatrix} :=\begin{pmatrix} p(c_+)& p(c_-)\\ 1-p(c_+)& 1-p(c_-)\end{pmatrix} 
\begin{pmatrix}\phi(c_+)\\ \phi(c_-)\end{pmatrix} 
\end{equation}
where $p(c_\pm)$ are the local transition probabilities corresponding to fixed configurations $c_\pm$ on three neighboring minimal double cones on the Cauchy surface $\CS_{t-\fel}$ and with configuration value $+1$ on the fourth minimal double cone in their (future) domain of dependence. Hence, the state extension backward in time, that is from the Cauchy surface algebra $\OA_t$ to the Cauchy surface algebra $\OA_{t-\fel}$ is defined uniquely by the inverse mappings
\begin{equation}
\frac{1}{p(c_+)-p(c_-)}
\begin{pmatrix} 1-p(c_-)& -p(c_-)\\ -1+p(c_+)& p(c_+)\end{pmatrix}
\begin{pmatrix}\phi(c^+)\\ \phi(c^-)\end{pmatrix} = \begin{pmatrix} p(c_+)& p(c_-)\\ 1-p(c_+)& 1-p(c_-)\end{pmatrix}^{-1} \begin{pmatrix}\phi(c^+)\\ \phi(c^-)\end{pmatrix} =:
\begin{pmatrix}\phi(c_+)\\ \phi(c_-)\end{pmatrix}\nonumber
\end{equation}
iff the four matrices are invertible, that is $p(c_+)\not= p(c_-)$ for the four possible choices of configurations of two neighboring minimal double cones in $S^m_t$. However, one has to ensure also the inequalities $0\leq \phi(c_+),\phi(c_-)\leq 1$, which in case of $\phi(c^+)+\phi(c^-)>0$ lead to restrictions for the ratio $\rho(c):=\phi(c^+)/\phi(c^-)$:
\begin{eqnarray}
p(c_+)&\geq& (1-p(c_+))\rho(c), (1-p(c_-))\rho(c)\geq p(c_+),
   \quad p(c_+)>p(c_-),\nonumber\\
p(c_-)&\geq& (1-p(c_+))\rho(c), (1-p(c_-))\rho(c)\geq p(c_),
   \quad p(c_-)>p(c_+).
\end{eqnarray}

Forgetting the difficulties of state extensions backward in time once a state $\phi_s\colon\OA_s\to\CO$ on the Cauchy surface subalgebra $\OA_s$ of the causal stochastic Ising model is given then the eight determining local transition probabilities $\{ p(c_\pm) \}$ as conditional probabilities give rise to the extension of $\phi_s$ to a state on time interval quasilocal algebras $\OA(s.t), t>s$. Having performed this extension Bell's local causality will hold in the time interval quasilocal algebras for any values of the eight determining transition probabilities due to Proposition \ref{BLC_stoch_LPC}.

Having established the validity of local causality in LPTs with local primitive causality and in stochastic LCTs without local primitive causality, in the next section we will review how Bell's notion of local causality relates to the Common Cause Principle and the Bell inequalities.

\section{Local causality, Common Cause Principle and the Bell inequalities} 

Local causality is closely related to Reichenbach's (1956) Common Cause Principle. The \textit{Common Cause Principle} (CCP) states that if there is a correlation between two events $A$ and $B$ and there is no direct causal (or logical) connection between the correlating events, then there always exists a common cause $C$ of the correlation. Reichenbach's original definition is formulated in a purely classical probabilistic setting lacking any spatiotemporal considerations; however, it can readily be generalized to the LPT framework. (For the steps of the generalization see (Rédei 1997, 1998), (Rédei and Summers 2002, 2007), (Hofer-Szabó and Vecsernyés 2012, 2013) and (Hofer-Szabó, Rédei and Szabó 2013).) 

Let $\{\vNA(V),V\in\SDK\}$ be a net representing a LPT. Let $A \in\vNA(V_A)$ and $B\in\vNA(V_B)$ be two events (projections) supported in spacelike separated regions $V_A, V_B\in\SDK$, which correlate in a locally normal and faithful state $\phi$. The common cause of a the correlation is an event $C$ which (together with its complement) screens off the correlating events from one another, and which is localized in the causal past of $A$ and $B$. For the precise choice of this past one has (at least) three options. One can localize $C$ either (i) in the \textit{union} or (ii) in the \textit{intersection} of the causal past of the regions $V_A$ and $V_B$; or (iii) more restrictively, in the spacetime region which lies in the intersection of causal pasts of \textit{every} point of $V_A \cup V_B$, formally $\cap_{x \in V_A \cup V_B}\, J_-(x)$; see (Rédei, Summers 2007). 
We will refer to the above three pasts in turn as the \textit{weak past}, \textit{common past}, and \textit{strong past} of $A$ and $B$, respectively. 

Now, we can define various CCPs in a LPT:
\begin{D}\label{CCP} A LPT represented by a net $\{\vNA(V),V\in\SDK\}$ is said to satisfy the \textit{(Weak/Strong) CCP}, if for any pair $A \in\vNA(V_A)$ and $B\in\vNA(V_B)$ of projections supported in spacelike separated regions $V_A, V_B\in\SDK$ and for every locally faithful state $\phi$ establishing a correlation between $A$ and $B$, there exists a nontrivial common cause system that is a set of mutually orthogonal projections $\{ C_k \}_{k\in K}\subset \vNA(V_C), V_C\in\SDK$ localized in the (weak/strong) common past of $V_A$ and $V_B$, which decompose the unit and satisfy 
\begin{equation}\label{CC}
\phi_{C_k}(AB)= \phi_{C_k}(A)\phi_{C_k}(B),\quad k\in K,
\end{equation}
where the state $\phi_{C_k}$ is given by (\ref{selectphi}).

A common cause is called \textit{trivial} if $C_k\leq X$ with $X=A,A^\perp, B$ or $B^\perp$ for all $k\in K$. If $C_k$ commutes with both $A$ and $B$ for all $k\in K$, then we call it a {\em commuting} common cause system, otherwise a {\em noncommuting} one, and the appropriate CCP a \textit{Commutative/Noncommutative CCP}.
\end{D}
Trivial common cause systems provide solutions of (\ref{CC}) independently of the state $\phi$. Therefore they are considered as purely `kinematic' or `algebraic' solutions that are insensitive to the actual physical environment provided by a particular state $\phi$. If  at least one of the algebras $\vNA(V_A)$ and $\vNA(V_B)$ is finite dimensional, then even a more trivial common cause system can be given which is not sensitive even to the given algebra elements $A$ and $B$. Namely, any decomposition of the unit into minimal projections of the corresponding finite dimensional algebra\footnote{Of course the cardinality $\vert K\vert$ of these (commuting or noncommuting) common cause systems is uniquely determined by the finite dimensional algebra: $\vert K\vert=\sum_r n_r$ if the finite dimensional algebra is isomorphic to finite direct sum of full matrix algebras, $\oplus_r M_{n_r}$.}, i.e. any \textit{maximal} (atomic) decomposition of the unit, provides a weak common cause system 
 solution of (\ref{CC}) irrespectively 
of the chosen events in $\vNA(V_A)$ and $\vNA(V_B)$, and irrespectively of the correlating state $\phi$ on them (Cavalcanti and Lal, 2013). 
Therefore these trivial, maximal size solutions reflect more the structure of the underlying finite dimensional local algebras, $\vNA(V_A)$ or $\vNA(V_B)$ or both, which contain them. For example, in this case $\phi_{C_k}, k\in K$ become always pure states by restriction to the corresponding finite dimensional algebra. Since $C_k, k\in K$ are spacelike separated to the other local algebra, (\ref{CC}) should hold in a locally causal theory for any choice of $A\in\vNA(V_A),B \in \vNA(V_B)$ and any locally faithful state $\phi$ on the quasilocal observables according to Definition 1.

To reveal the similarities and the differences between Bell's local causality and the CCPs we note that the core mathematical requirement of both properties is the screening-off conditions (\ref{BLC_screening}) or equivalently (\ref{CC}). However, the subjects of these conditions are very different: In the first case the screening-off should hold for all pairs of algebra elements supported in the spacelike regions $V_A,V_B\in\SDK$. On the contrary, different common cause systems are not only allowed for different triples $(A,B,\phi)$ but also a nontrivial dependence is expected on physical grounds. Moreover, in case of local causality the screening-off condition (\ref{BLC_screening}) is required for \textit{every} atomic event (satisfying certain localization conditions). 
In case of the CCP the screening-off condition (\ref{CC}) should be satisfied only by a \textit{single subset} of events, by a decomposition of unit,  which, apart from the `kinematic' maximal size solution, is typically not given by atomic events.

However, there is an exciting similarity: there exist derivations of the Bell inequalities from both conditions (together with some additional requirements). In (Hofer-Szabó and Vecsernyés, 2013b, Proposition 2) we have proven a proposition which clarifies the relation between the CCPs and the Bell inequalities. It asserts that the Bell inequalities can be derived from the existence of a common cause system for a set of correlations \textit{if} common causes are understood as \textit{commuting} common causes. However, if we also allow for \textit{non}commuting common causes, the Bell inequalities can be derived only for another state which is \textit{not} identical to the original one. And indeed in (Hofer-Szabó and Vecsernyés, 2013a,b) a noncommuting common cause was constructed for a set of correlations violating the Clauser--Horne inequality. Moreover, this common cause was localized in the strong past of the correlating events. 

Now, an analogous proposition holds for the relation between local causality and the Bell inequalities. We assert here only the proposition without the proof since the proof is step-by-step the same as that of the proposition mentioned above.
\begin{Prop}\label{PqCH} 
Let $\{\vNA(V),V\in\SDK\}$ be a locally causal LPT with atomic (type I) local von Neumann algebras. Let $A_1, A_2  \in \mathcal A(V_A)$ and  $B_1, B_2 \in \mathcal A(V_B)$  be four projections localized in spacelike separated spacetime regions $V_A$ and $V_B$, respectively, which pairwise correlate in the locally faithful state $\phi$ that is 
\begin{eqnarray} \label{qcorrs}
\phi(A_mB_n) &\neq& \phi(A_m)\, \phi(B_n)
\end{eqnarray}
for any $m,n=1,2$. Let $V_C\in\SDK$ be a region satisfying requirements $L_1$, $L_2$ and $L^Q_3/L^C_3$ in Definition \ref{BellLC} of local causality and let $\{ C_k \}_{k\in K}\subset \vNA(V_C)$ be a maximal partition of unit containing mutually orthogonal \textit{atomic} projections. Then the Clauser--Horne inequality 
\begin{eqnarray}\label{qCH_mn_noncomm} 
-1 \leqslant (\phi \circ \OP_{\{C_k\}}) (A_1 B_1 + A_1 B_2 + A_2 B_1 - A_2 B_2 - A_1 - B_1) \leqslant 0.
\end{eqnarray}
holds for the state $\phi \circ \OP_{\{C_k\}}$. If $\left\{ C_k \right\}$ commutes with $A_1$, $A_2$, $B_1$ and $B_2$, then the Clauser--Horne inequality holds for the original state $\phi$: 
\begin{eqnarray}\label{qCH_mn} 
-1 \leqslant \phi (A_1 B_1 + A_1 B_2 + A_2 B_1 - A_2 B_2 - A_1 - B_1) \leqslant 0.
\end{eqnarray}
\end{Prop}
\vspace{0.2in}

\noindent 
The moral is the same as in the case of the CCPs: the Bell inequalities can be derived in a locally causal LPT only for a \textit{modified} state $\phi \circ \OP_{\{C_k\}}$ in general. It can be derived for the \textit{original} state $\phi$ \textit{if} the set of atomic projections $\{ C_k \}$ localized in $V_C$ commutes with $A_1$, $A_2$, $B_1$ and $B_2$. Clearly, if the LPT is classical,  the elements taken from any local algebra will commute, therefore Bell inequalities hold for the original state $\phi$ in LCTs. However, going over to locally causal LQTs, commutation of $\{ C_k \}$ with the correlating events is not guaranteed. If $V_C$ is spatially separated from $V_B$ (ensured by requirement $L^Q_3$ but not $L^C_3$), then $\{ C_k \}$ will commute with $B_1$ and $B_2$ due to microcausality, hence (\ref{BLC_screening}) will be satisfied, even if the $B_1$ and $B_2$ do not commute. However, in case of local primitive causality one cannot pick a maximal partition of unit 
$\{ C_k \}$ in $\vNA(V_C)$ (which is needed for the states $\phi_{C_k}$ 
to be pure on $\vNA(V_C)$) such that $\{ C_k \}$ commutes also with projections $A_1$ and $A_2$, if $[A_1,A_2]\neq 0$. Namely, $\vNA(V_A)\subset\vNA(V_C'')=\vNA(V_C)$ due to isotony and local primitive causality, and the image $\OP_{\{C_k\}}(\vNA(V_C))$ is a maximal \textit{abelian} subalgebra of $\vNA(V_C)$ containing exactly those elements that commute with $\{C_k\}$. Hence, in order to commute with $\{ C_k \}$, both $A_1$ and $A_2$ should be contained in $\OP_{\{C_k\}}(\vNA(V_C))$, which cannot be the case, if $[A_1,A_2]\neq 0$.

The conclusion is that in case of noncommuting projections $A_1$ and $A_2$ the theorem of total probability, $\sum_k \phi(C_kA_mC_k) = \phi(A_m)$, will not hold for the original state\footnote{It holds only for the state $\phi_{\{C_k\}}$ for which $\phi_{\{C_k\}}(A_m)\not=\phi(A_m)$ at least for one of the projections $A_1$ and $A_2$.} $\phi$ at least for one of the projections $A_1$ and $A_2$. This fact blocks the derivation of Bell inequalities for the original state $\phi$. (For the details see (Hofer-Szabó and Vecsernyés, 2013b, p. 410).) In short, the Bell inequalities can be derived in a locally primitive causal LQT with atomic von Neumann algebras, hence in a locally causal LQT, only if the projections supported on both of the correlating regions commute.\footnote{Although local causality is formulated asymmetrically for $A$ and $B$, this asymmetric prescription should hold for the case of interchanged role with respect to $\OP$, too.}

Coming back to the question posed at the end of the previous section, namely how a local \textit{quantum} theory can be locally causal in the face of the Bell inequalities, we already know the answer: the Bell inequalities can be derived from local causality if the 'beables' of the local theory are represented by \textit{commutative} local algebras. This fact is completely analogous with the relation shown in (Hofer-Szabó and Vecsernyés, 2013b): Bell inequalities can be derived from a (joint, nonconpiratorial, local) common cause system if it is a \textit{commuting} common cause system. Thus, both common causal explanation and local causality are more general notions than what is captured by the Bell inequalities.

\section{Summary}

In this paper we aimed to give a clear-cut definition of Bell's notion of local causality. To this end, first we unfolded a framework, called local physical theory, which integrates probabilistic and spatiotemporal concepts in a common conceptual schema. We have clarified how primitive causality and local primitive causality lead to deterministic and causal dynamics, respectively. We have introduced the notion of causal Markov process with independent transition probabilities with respect to spacelike separation and showed that they lead to a causal stochastic dynamics interpretation of the state extension procedure in LCTs without primitive causality. 
Having formulated Bell's local causality within the framework of LPTs we have given sufficient conditions for a LPT to be locally causal: 1. local primitive causality holds and the local von Neumann algebras are atomic, 2. primitive causality does not hold but the state on the quasilocal algebra arises from the mentioned causal stochastic process and the local von Neumann algebras are finite dimensional. 
We have constructed an explicit model for the latter case, called stochastic causal Ising model. We compared Bell's local causality with the various Common Cause Principles and related both to the Bell inequalities. We found a nice parallelism: Bell inequalities cannot be derived neither from local causality nor from a common cause unless the local physical theory is classical or the common cause is commuting, respectively.
\vspace{0.2in}

\noindent
{\bf Acknowledgements.}  This work has been supported by the Hungarian Scientific Research Fund, OTKA K-100715 and OTKA K-108384.

\section*{References} 
\begin{small}
\begin{list} 
{ }{\setlength{\itemindent}{-15pt}
\setlength{\leftmargin}{15pt}}

\item J.S. Bell, ''Beables for quantum field theory,'' TH-2053-CERN, presented at the Sixth GIFT Seminar, Jaca, 2–7 June (1975); reprinted in (Bell, 2004, 52-62).

\item J.S. Bell, ''EPR correlations and EPW distributions,'' in: {\it New Techniques and Ideas in Quantum Measurement Theory}, New York Academy of Sciences, (1986); reprinted in (Bell, 2004, 196-200).

\item J.S. Bell, ''La nouvelle cuisine,'' in: J. Sarlemijn and P. Kroes (eds.), {\it Between Science and Technology}, Elsevier, (1990); reprinted in (Bell, 2004, 232-248).

\item J.S. Bell, \textit{Speakable and Unspeakable in Quantum Mechanics}, (Cambridge: Cambridge University Press, 2004).

\item R. Brunetti and K. Fredenhagen, ''Quantum Field Theory on Curved Backgrounds,'' in C. B\"ar and K. Fredenhagen (eds.), \textit{Quantum Field Theory on Curved Spacetimes, Concepts and Mathematical Foundations}, Lecture Notes in Physics 786, (Berlin: Springer-Verlag, 2009).

\item J. Butterfield, ''Vacuum correlations and outcome independence in algebraic quantum field theory'' in D. Greenberger and A. Zeilinger (eds.), \textit{Fundamental Problems in Quantum Theory, Annals of the New York Academy of Sciences, Proceedings of a conference in honour of John Wheeler}, 768-785 (1995).

\item J. Butterfield, ''Stochastic Einstein Locality Revisited,''  \textit{Brit. J. Phil. Sci.}, \textbf{58},  805-867, (2007).

\item E. G. Cavalcanti and R. Lal, ``On modifications of Reichenbach's principle of common cause in light of Bell's theorems,'' http://arxiv.org/abs/1311.6852 (2013).

\item R. Clifton and H. Halvorson, ''Entanglement and Open Systems in
Algebraic Quantum Field Theory,'' \textit{Stud. Hist. Phil. Mod. Phys.},  \textbf{32 (1)}, 1–31 (2001).

\item L. Diósi, ``Models for universal reduction of macroscopic quantum fluctuations'' \textit{Phys. Rev. A}, \textbf{40}, 1165 (1989).

\item J. Earmen, ''No superluminal propagation for classical relativistic quantum fields,'' (manuscript), (2014).

\item J. Earman and G. Valente,  ''Relativistic causality in algebraic quantum field theory,'' \textit{Int. Stud. Phil. Sci.}, (forthcoming) (2014).

\item R. Geroch, ''Faster than light?,'' http://arxiv.org/abs/1005.1614 (2010).

\item G. C. Ghirardi, A., Rimini and T. Weber, ''A Model for a Unified Quantum Description of Macroscopic and Microscopic Systems,'' in L. Accardi et al. (eds.) \textit{Quantum Probability and Applications} (Berlin: Springer-Verlag, 1985).

\item R. Haag, \textit{Local quantum physics}, (Heidelberg: Springer Verlag, 1992).

\item H. Halvorson, ''Algebraic quantum field theory,'' in J. Butterfield, J. Earman (eds.), \textit{Philosophy of Physics, Vol. I}, Elsevier, Amsterdam, 731-922 (2007).

\item J. Henson, ''Comparing causality principles,'' \textit{Stud. Hist. Phil. Mod. Phys.}, \textbf{36}, 519-543 (2005).

\item J. Henson, ''Confounding causality principles: Comment on Rédei and San Pedro's ``Distinguishing causality principles'','' \textit{Stud. Hist. Phil. Mod. Phys.}, \textbf{44}, 17-19 (2013a).

\item J. Henson, ''Non-separability does not relieve the problem of Bell's theorem,'' \textit{Found. Phys.}, \textbf{43}, 1008-1038 (2013b).

\item G. Hofer-Szabó, M. Rédei and L. E. Szabó, \textit{The Principle of the Common Cause}, (Cambridge: Cambridge University Press, 2013).

\item G. Hofer-Szab\'o and P. Vecserny\'es, ''Reichenbach's Common Cause Principle in AQFT with locally finite degrees of freedom,'' \textit{Found. Phys.}, \textbf{42}, 241-255 (2012a).

\item G. Hofer-Szab\'o and P. Vecserny\'es, ''Noncommuting local common causes for correlations violating the Clauser--Horne inequality,'' \textit{J. Math. Phys.}, \textbf{53}, 12230 (2012b).

\item G. Hofer-Szab\'o and P. Vecserny\'es, ''Noncommutative Common Cause Principles in AQFT,'' \textit{J. Math. Phys.}, \textbf{54}, 042301 (2013a).

\item G. Hofer-Szab\'o and P. Vecserny\'es, ''Bell inequality and common causal explanation in algebraic quantum field theory,'' \textit{Stud. Hist. Phil. Mod. Phys.},  \textbf{44 (4)}, 404–416 (2013b).

\item G. Hofer-Szab\'o and P. Vecsernyés, ''Bell's local causality for philosophers,'' (in preparation) (2014).

\item G. Hofer-Szab\'o, ''Local causality and complete specification: a reply to Seevinck and Uffink,'' submitted to EPSA 2013 Proceedings (2015a).

\item G. Hofer-Szab\'o, ''Relating Bell's local causality to the Causal Markov Condition,'' submitted to BCAP 2013 Proceedings (2015b).

\item F. Károlyházy,  ''Gravitation and quantum mechanics of macroscopic objects,'' \textit{Nuovo Cimento A}, \textbf{42}, 390 (1966).

\item G. Lüders: ''Über die Zustandsänderung durch den Messprozess,'' \textit{Annalen der Physik}, \textbf{443}, 322 (1950).

\item C. Mann and R. Crease, ''John Bell, Particle Physicist,'' (Interview), \textit{Omni}, \textbf{10/8}, 84-92 (1988).

\item T. Norsen, ''Local causality and Completeness: Bell vs. Jarrett,'' \textit{Found. Phys.}, \textbf{39}, 273 (2009).

\item T. Norsen, ''J.S. Bell's concept of local causality,'' \textit{Am. J. Phys}, \textbf{79}, 12, (2011).

\item F. Pf\"affle, ''Lorentzian manifolds,''  in C. B\"ar and K. Fredenhagen (eds.), \textit{Quantum Field Theory on Curved Spacetimes, Concepts and Mathematical Foundations}, Lecture Notes in Physics 786, (Berlin: Springer-Verlag, 2009).

\item M. R\'edei, ''Reichenbach's Common Cause Principle and quantum field theory,'' \textit{Found. Phys.}, \textbf{27}, 1309-1321 (1997).

\item M. Rédei and I. San Pedro, ''Distinguishing causality principles,'' \textit{Stud. Hist. Phil. Mod. Phys.}, \textbf{43}, 84-89 (2012). 

\item M. R\'edei and J. S. Summers, ''Local primitive causality and the Common Cause Principle in quantum field theory,'' \textit{Found. Phys.}, \textbf{32}, 335-355 (2002).

\item H. Reichenbach, {\it The Direction of Time}, (University of California Press, Los Angeles, 1956).

\item M. P. Seevinck and J. Uffink, ''Not throwing our the baby with the bathwater: Bell's condition of local causality mathematically 'sharp and clean', '' in: Dieks, D.; Gonzalez, W.J.; Hartmann, S.; Uebel, Th.; Weber, M. (eds.) \textit{Explanation, Prediction, and Confirmation The Philosophy of Science in a European Perspective}, Volume 2, 425-450 (2011).

\item S. Schlieder, ''Einige Bemerkungen \"uber Projektionsoperatoren (Konsequenzen eines Theorems von Borchers,'' \textit{Comm. Math. Phys.}, \textbf{13},  216-225 (1969).

\item A. Shimony, ''Events and processes in the quantum world,'' in Penrose, R. and Isham, C. (eds.), \textit{Quantum concepts in space and time}, 182–203. (Oxford: Oxford University Press, 1986.)

\item S. J. Summers and R. Werner, ''Bell's inequalities and quantum field theory, I: General setting,'' \textit{J. Math. Phys.}, \textbf{28}, 2440-2447 (1987a).

\item S. J. Summers and R. Werner, ''Bell's inequalities and quantum field theory, II: Bell's inequalities are maximally violated in the vacuum,'' \textit{J. Math. Phys.}, \textbf{28},  2448-2456 (1987b).

\item S. J. Summers and R. Werner, ''Maximal violation of Bell's inequalities for algebras of observables in tangent spacetime regions,'' \textit{Ann. Inst. Henri Poincar\'e -- Phys. Th\'eor.}, \textbf{49}, 215-243 (1988).

\item R. Werner, ''Local preparability of states and the split property in quantum field theory,'' \textit{Lett. Math. Phys.}, \textbf{13},  325-329 (1987).

\end{list}
\end{small}
\end{document}